\newcommand{\be}{\begin{equation}}
\newcommand{\ee}{\end{equation}}
\newcommand{\bea}{\begin{eqnarray}}
\newcommand{\eea}{\end{eqnarray}}

\documentclass[a4paper,11pt,onecolumn]{article}
\pdfoutput=1
\usepackage{jheppub}
\usepackage[T1]{fontenc}
\usepackage{graphicx}% Include figure files
\usepackage{dcolumn}% Align table columns on decimal point
\usepackage{bm}% bold math
\usepackage{amsmath}
\usepackage{amssymb}
\usepackage{slashed}
\usepackage{accents}
\def\d{d\kern-.8 ex\vrule height 1.3 ex depth-1.24 ex width .7 ex \kern .15 ex}
\def\D{D\kern-1.7 ex\vrule height .87 ex depth-.8 ex width .7 ex \kern .95 ex}

\title{Weak chaos and mixed dynamics in the string S-matrix}

\author[a,b]{Nikola Savi\'c}
\author[b]{and Mihailo \v{C}ubrovi\'c}
\affiliation[a]{Institut de Physique Theorique, Saclay, Gif-sur-Ivette, France}
\affiliation[b]{Institute of Physics Belgrade, University of Belgrade, Pregrevica 118, 11080 Belgrade, Serbia}
\emailAdd{nikola.savic@ipht.fr}
\emailAdd{cubrovic@ipb.ac.rs}

\date{\today}

\abstract{We investigate chaotic dynamics in tree-level S-matrices describing the scattering of tachyons, photons and gravitons on highly excited open and closed bosonic strings, motivated by the string/black hole complementarity. The eigenphase spacing distribution and other indicators of quantum chaotic scattering suggest that the dynamics is only weakly chaotic, consisting of both regular/Poisson and chaotic/Wigner-Dyson processes. Only for special values of momenta and (for photon scattering) scattering angles do we find strong chaos of random matrix type. These special values correspond to a crossover between two regimes of scattering, dominated by short versus long partitions of the total occupation number of the highly excited string; they also maximize the information entropy of the S-matrix. The lack of strong chaos suggests that perturbative dynamics of highly excited strings can never describe the universal properties and maximal chaos of black hole horizons.}

\begin{document} 
\maketitle
\flushbottom

\section{Introduction}\label{secintro}

Chaos and dynamical phenomena have turned out to provide a fundamental insight into the structure and information content of black holes and gravity in general. Since the discovery of the black hole scrambling concept \cite{FastScramblers,FastScramblersHayden}, chaos bound and OTOC-ology \cite{MSSbound,Polchinski:2015cea} and on the other hand the questions of factorization and microscopic statistics stemming from the replica wormhole proposal in the context of the information paradox \cite{BHinfoAlmheiri:2019,BHinfoPenington:2019,BHinfoRMP}, it is becoming clear that nonlinear dynamics is an integral part of high-energy theory, as many questions in gravity and string theory are naturally formulated in terms of scrambling, thermalization and microscopic chaos. 

The time is now ripe to move beyond semiclassical horizons and black holes in general relativity. The same questions of scrambling, universal timescales and the like can be asked also for stringy black hole solutions \cite{David:2002wn} or for solutions proposed to replace black holes by horizonless stringy objects (microstate geometries and fuzzballs) \cite{Bena:2022rna}. With stringy corrections, things become much more difficult and less universal. One possible approach is the string/black hole (string/BH) complementarity paradigm \cite{Susskind:1993ws,Horowitz:1996BHstring,Horowitz:1997BHstring}: a highly excited string should look like a black hole in the weak coupling regime.

The idea of string/BH complementarity stems from the fact that at sufficiently high occupation numbers, the Schwarzschild radius of the HES becomes smaller than the string scale, hence the string should collapse into a black hole. To remind, the mass of a string $M_s$ and the mass of a black hole $M_\mathrm{BH}$ in $d+1$ spacetime dimension are
\be 
M_\mathrm{BH}\sim\frac{r_s^{d-2}}{G},~M_s\sim\frac{N}{\alpha'},
\ee
where $\alpha'$ is the string tension, $G$ is the Newton's constant and $N$ the occupation number (level). At the string/BH transition, we expect the string length scale to be $\ell_s=\sqrt{\alpha'}\sim r_s$; then a string of mass $M_s$ can become a black hole of mass $M_\mathrm{BH}=M_s\equiv M$. Equating the mass and length scales and taking into account that $G\sim g^2\alpha'$ where $g$ is the string coupling, we find the condition for the black hole description of the string:
\be 
Ng^4\sim (\alpha')^{d-3}\Rightarrow g_c\propto N^{-1/4}(\alpha')^{\frac{d-3}{4}}.
\ee
Therefore, when the total occupation number $N$ is large the string will approximately describe a black hole already at small $g$, i.e. in the perturbative regime. This is the motivation behind the recent works on the scattering amplitudes of highly excited strings (HES). By HES we mean simply a string with $N\gg 1$, and scattering amplitudes of the form $\mathrm{HES}\to\mathrm{HES}+\mathrm{A}$ or $\mathrm{HES}+\mathrm{A}\to\mathrm{HES}+\mathrm{B}$ provide a stringy equivalent to the scrambling perspective for black holes, where we scatter a field on an AdS black hole, obtaining a time-disordered correlation function as a probe of chaos \cite{BlackButter,ShocksLocal,ShocksStringy}. In \cite{Gross:2021gsj} a general framework for computing the HES scattering amplitudes was formulated, based on the Del Giudice-Di Vecchia-Fubini (DDF) formalism \cite{DDFfirst,DDFcoherent,Skliros:2011si,Bianchi:2019ywdDDF}. The idea is to obtain the HES in a controlled way, by adding photonic excitations to a tachyon (vacuum of the bosonic string). In \cite{Rosenhaus:2021PRL,Firrotta:2022Photon,Hashimoto:2022bll,Firrotta:2023wem} the poles and zeros of the resulting amplitudes are studied within the framework of random matrix theory (RMT), also making use of a novel chaos indicator, the ratio of eigenvalue spacings, discussed in \cite{Bianchi:2022mhs,Bianchi:2023uby}.

We are thus dealing with (possibly chaotic) quantum scattering. The reader is perhaps better acquainted with the study of quantum chaos in closed systems, i.e. the study of quantum Hamiltonians \cite{Haake1991}, where strong chaos is successfully described by the RMT approach \cite{MehtaRMT}. To remind, quantum chaotic Hamiltonians of sufficiently large size approach Gaussian random matrices in their behavior, and can be described by the Gaussian ensemble statistics. In particular, the eigenvalue (i.e., eigenenergy) spacings obey the celebrated Wigner-Dyson distribution, and the system is chaotic in the RMT sense if the eigenvalue spacing histogram is well fit by the Wigner-Dyson curve.

A scattering problem is inherently different. Instead of a Hamiltonian, we deal with scattering amplitudes and the S-matrix which describes the probability amplitudes of all possible scattering outcomes. A difference of principle is that the scattered object ends up "at infinity", i.e. the system is not closed in the usual sense. For classical scattering, this means that exit curves behave in a sense like attractors in dissipative chaos, i.e. they tell us where the orbits end up when $t\to\infty$. A measure of chaos is then the very complex (usually fractal) dependence of the final state (e.g. scattering angle) on the initial conditions. Therefore, we might expect a similar fractal behavior for the dependence of amplitudes on the scattering angle. This approach was used in \cite{Hashimoto:2022bll}. Another approach is to look at the spacings of special points in the amplitudes and to test if these satisfy the RMT statistics, analogously to the eigenenergy spacings in Hamiltonian systems; this approach was taken in \cite{Rosenhaus:2020PRL,Rosenhaus:2021PRL,Firrotta:2022Photon,Firrotta:2023wem,Bianchi:2023uby}. The outcome of these works is very interesting and has uncovered several important ideas, e.g. the emergence of thermalization in \cite{Firrotta:2023wem} but concerning the chaos itself it is inconclusive: while there are clear signs of chaos some indicators, e.g. the spacing ratios in \cite{Bianchi:2023uby} deviate significantly from the RMT predictions, and the angular dependence of the amplitudes, while complex, is not fractal \cite{Hashimoto:2022bll}.

In this work we continue the exploration of chaos in HES scattering but attempt a more systematic approach, studying directly the S-matrix rather than individual amplitudes. We also introduce a novel element of "geometric" chaos which arises when we look at non-scalar scatterers, i.e. the process $\mathrm{HES}+\mathrm{photon}$ instead of $\mathrm{HES}+\mathrm{tachyon}$. For the S-matrix a rigorous generalization of the RMT approach exists \cite{EDoron_1992,Smilansky1992}: the role of eigenenergies is taken by the eigenphases, i.e. the phases of the (complex) eigenvalues of the S-matrix. Now the eigenphase spacings are expected to obey the Wigner-Dyson distribution. It will turn out that the study of the S-matrix as a whole reveals some additional surprises: we find clear signs of chaos only for special values of the angles and/or momenta, and the underlying mechanism is the competition between different ("short" and "long") partitions of the total occupation number $N$. This "combinatorial" approach to chaos will also reveal the existence of quasi-invariant states, which necessarily spoil the chaos -- in other words, despite finding strong chaos at special points in parameter space, we argue that there is always a regular component. All of the above holds for bosonic strings: we do not consider superstrings in this work so from now on it is understood that "string" means bosonic string.

Recently the work \cite{Das:2023cdn} has appeared which also studies the properties of the S-matrix and likewise sees a crossover between "short" and "long" partitions as dominant in the scattering (though their terminology is different). There is therefore some overlap of our work and \cite{Das:2023cdn} but our results are mostly complementary: in \cite{Das:2023cdn} the influence of the polarization of the DDF photons is studied, typicality of states is probed and in addition important indications of eigenstate thermalization are found, whereas we study in more detail the dynamics itself and introduce the photon (instead of tachyon) scattering. We warmly recommend the reader to study \cite{Das:2023cdn} in addition to our paper.

With some hindsight, we can say that our results indicate that HES scattering is never uniformly chaotic and therefore does not directly provide a look at black holes in the stringy regime; the naive hope that the string/BH complementarity can be seen directly in the HES S-matrix is thus invalid. But this conclusion is also useful for future work, and the HES S-matrix is in itself worth studying, as it shows some novel phenomena of quantum chaotic scattering.

The plan of the paper is the following. In Section \ref{seches} we briefly recapitulate the construction of the HES in the DDF formalism and the calculation of the HES-tachyon amplitude; then we compute the HES-photon amplitude which was never studied so far. In Section \ref{secchaos} we introduce the tools and ideas for studying the S-matrix dynamics: eigenphase statistics and the combinatorics of partitions/states. Section \ref{secnum} brings the results of the analysis described in Section \ref{secchaos}, and in Section \ref{secconc} we discuss the implications of our findings and directions of further work.

\section{Highly excited strings and their scattering amplitudes}\label{seches}

In this section we set the stage: we construct the highly excited string states and then write the tree-level scattering amplitude for a few simple $2\to 2$ processes, scattering a tachyon $\mathrm{t}$ or a photon $\gamma$ off a HES:
\bea 
\mathrm{HES}+\mathrm{t}&\longrightarrow&\mathrm{HES'}+\mathrm{t'}\label{process}\\
\mathrm{HES}+\gamma&\longrightarrow&\mathrm{HES'}+\gamma'.\label{processgamma}
%\mathrm{HES}+\mathrm{g}&\longrightarrow&\mathrm{HES'}+\mathrm{g'}.\label{processgrav}
\eea
For open strings we closely follow the formalism of \cite{Skliros:2011si,Gross:2021gsj} and specifically the calculations of \cite{Hashimoto:2022bll}. For closed strings we construct the amplitude exploiting the KLT relations \cite{KLT,Sondergaard:2011ivKLT,Stieberger:2014hbaKLT} and check the result by direct integration on the worldsheet. The KLT duality applied to Eqs.~(\ref{process}) and (\ref{processgamma}) will yield the processes with closed HES (cHES) and tachyon or graviton respectively:
\bea 
\mathrm{cHES}+\mathrm{t}&\longrightarrow&\mathrm{cHES'}+\mathrm{t'}\label{processc}\\
\mathrm{cHES}+\mathrm{g}&\longrightarrow&\mathrm{cHES'}+\mathrm{g'}.\label{processgrav}
\eea
In the rest of the paper we do not differentiate between open and closed HES in notation, i.e. we use HES (rather than cHES) for both cases, as the open/closed nature of the string will always be stated explicitly. Of course, we do differentiate between a photon $\gamma$ and a graviton $\mathrm{g}$ as the two have different spins. We will thus always call the process in Eq.~(\ref{processgamma}) the HES-photon scattering and the process in Eq.~(\ref{processgrav}) will be called the HES-graviton scattering.

\subsection{Open HES - tachyon amplitudes}\label{sechesopen}

\begin{figure}
        \centering
        \includegraphics[width=0.7\linewidth]{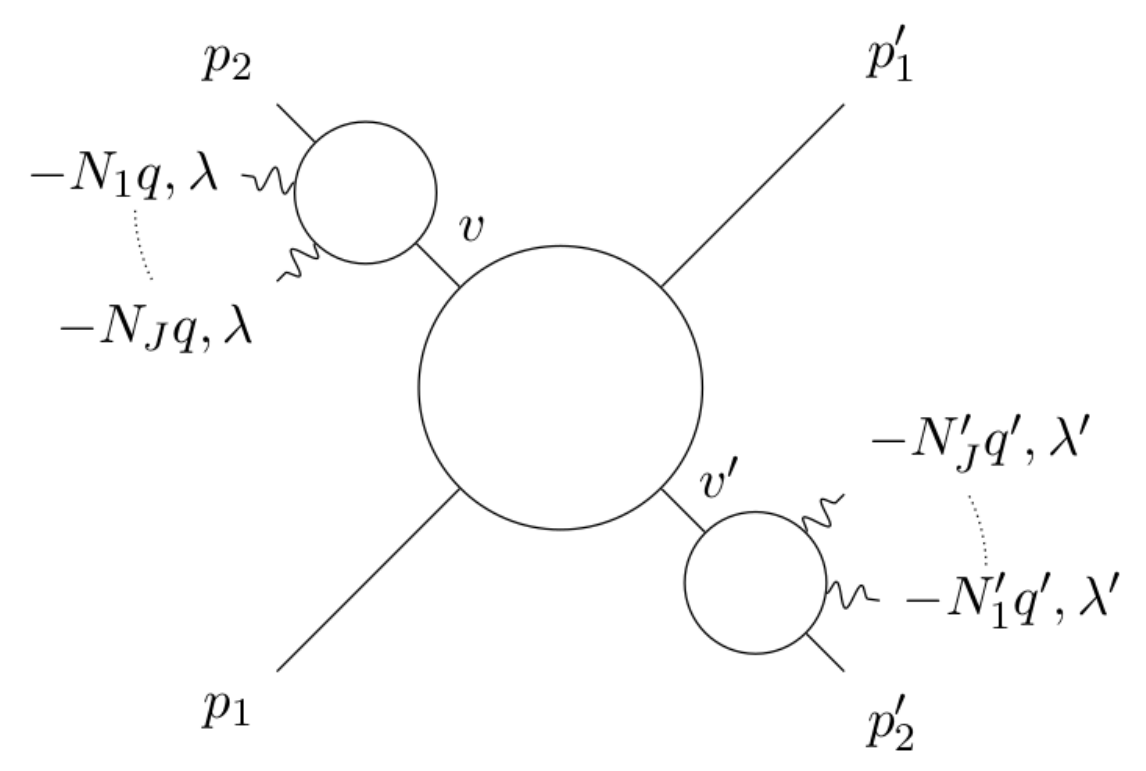}
        \caption{String amplitude for the HES-tachyon scattering process (\ref{process}). The tachyon labeled by its momentum $p_2$ absorbs $J$ photons with momenta and polarizations $\{-N_k q,\lambda\}$ which (after picking out appropriate poles) results in a $\mathrm{HES}$ state labeled by $v$ (similarly for the $\mathrm{HES'}$). The two $\mathrm{HES}$ states interact with tachyons labeled by $p_1$ and $p'_1$. Adapted from \cite{Hashimoto:2022bll}.} 
        \label{kojifig}
\end{figure}

The HES represents the state with a large number of excitations $\{n_k\}$, or equivalently with a high level $N = \sum_{k=1}^\infty n_k$, where $k$ labels the modes and $n_k$ is the occupation number of the $k$-th mode. Following the DDF formalism \cite{DDFfirst,Skliros:2011si,Bianchi:2019ywdDDF}, a convenient way to think of the HES state is as a state created in a process in which the tachyon absorbs $J$ photons (we will call them DDF photons) with momenta $q_a$ and polarizations $\lambda_a$ ($a=1,\ldots J$), one by one, as in Fig.~\ref{kojifig}. In the lightcone quantization the general HES state then has the form \cite{Skliros:2011si}:
\be
\vert\mathrm{HES}\rangle\propto\xi^{i_1\ldots i_J} P\left(\partial X,\left(\partial X\right)^2,\ldots\left(\partial X\right)^N\right)_{i_1\ldots i_J} \vert 0,p\rangle
\ee
where $\xi^{i_1\ldots i_J}$ is the polarization tensor, $P$ is a polynomial of degree $N$ over the derivatives of the string coordinates $X^\mu$ and $\vert 0,p\rangle$ is the tachyon state (ground state of the bosonic string) with momentum $p$. Hence, the physical process in Eq.~(\ref{process}) can be described by writing it formally as
\be
\mathrm{tachyon(1)} + \left(\mathrm{tachyon\left(2\right)} + J~\mathrm{photons}\right) \longrightarrow \mathrm{tachyon(1')} + \left(\mathrm{tachyon\left(2'\right)} + J'~\mathrm{photons}\right),\label{processddf}
\ee
and then picking out the poles so that the $\left(\mathrm{tachyon} + J~\mathrm{photons}\right)$ part of the amplitude creates an intermediate on-shell HES state, as shown schematically in Fig.~\ref{kojifig}.

For this procedure to correctly describe the absorption of $J$ photons by the tachyon one by one, it is necessary to remove the terms which couple the photons to each other. These are proportional to $\lambda_a\cdot\lambda_b$, so we must take the polarizations of different photons to be orthogonal to each other. For the sake of simplicity, we achieve this by making two special choices, again following \cite{Hashimoto:2022bll}:\footnote{These special choices essentially limit the polarization tensor $\xi^{i_1\ldots i_J}$ to a subset of all possible values. We have not explored more general choices; a more detailed discussion of the influence of the DDF photon polarization can be found in \cite{Das:2023cdn}.}
\begin{enumerate}
\item We take all photons to have the same polarization $\lambda_a\equiv\lambda$ such that $\lambda\cdot\lambda = 0$.
\item The momenta of DDF photons are taken to be equal to $q_a = -N_a q$ ($a=1,\ldots J$) where $N_a$ is a positive integer. Now $q$ must satisfy the condition $q \cdot \lambda = 0$ (this essentially states that all polarizations are transverse, as they have to be for photons).
\end{enumerate}
For the $\mathrm{HES'}$ state, on the right-hand side of Eq.~(\ref{process}), we similarly take $\lambda'\cdot\lambda'=0$ and $q'_b=-N_b' q'$. To find the tree-level amplitude we use the tachyon vertex operator $V_\mathrm{t}(z,p) = :e^{i p X(z)}:$, then we employ a computational trick to replace the photon vertex operators $V_\mathrm{p}(z,p)$ by $:e^{i\zeta\partial X+ikX}:$ and keep only the part linear in the polarization $\zeta$.

When the HES state is constructed as described above, we can express the tree-level amplitude as the path integral over the product of the vertex operators:
\be
\mathcal{A} = \int\frac{\mathcal{D}Xe^{-S_\mathrm{P}}}{\mathrm{Vol}\left(\mathrm{SL}\left(2,\mathbb{R}\right)\right)}\int \prod_i dw_i V_t(w_i,p_i) \prod_{a=1}^J dz_a V_p(z_a,-N_a q,\lambda) \prod_{b=1}^{J'} dz'_b V_p(z'_b,-N'_b q,\lambda)\label{amptachyon}
\ee
where $i\in \{ 1,2,1',2' \}$ runs over the tachyons ($1$ and $1'$ are physical tachyons and $2$ and $2'$ are DDF tachyons), $a\in \{ 1,\ldots,J \}$ runs over the photons in $\mathrm{HES}$, $b \in \{ 1,\ldots,J' \}$ runs over the photons in $\mathrm{HES'}$ and the integration variables $z$ and $w$ run over the worldsheet. The action in the path integral is the usual Polyakov action $S_\mathrm{P} = -\frac{1}{4 \pi \alpha'} \int_z dz (\partial X)^2$. For open string calculations we set $\alpha' = \frac{1}{2}$, while for closed strings in subsection \ref{sechesclosed} we take $\alpha'=2$. Next, still following \cite{Hashimoto:2022bll}, we make two more special choices to further simplify calculations:
\begin{enumerate}
\item The polarizations of DDF photons in HES and HES' satisfy $\lambda'\propto\lambda$.
\item The momenta of DDF photons in HES and HES' satisfy $q'\propto q$.\footnote{This choice was introduced in \cite{Bianchi:2019ywdDDF} and put to use also in \cite{Addazi:2020obs,Aldi:2020qfu}.}
\end{enumerate}
Using the above simplifications and performing the contractions we obtain:
\bea
    \mathcal{A} &=& \frac{1}{\mathrm{Vol}\left(\mathrm{SL}\left(2,\mathbb{R}\right)\right)}\int_{-\infty}^{\infty} \prod_{i} dw_i \int_{-\infty}^{\infty}\prod_{a=1}^J dz_a \int_{-\infty}^{\infty}\prod_{b=1}^{J'} dz'_b 
    \times\nonumber\\ &\times& \prod_{i < j} \vert w_i-w_j\vert^{p_i \cdot p_j}\prod_{a,i} \vert z_a - w_i\vert^{-\alpha_i} \sum_{i} \frac{-p_i \cdot \lambda}{w_i -z_a} \prod_{b,j} \vert z'_b - w_j\vert^{-\beta_j} \sum_{j} \frac{-p_j \cdot \lambda'}{w_j -z'_b},\label{openamp}
\eea
where we have expanded the integrand to linear order in photon polarizations, and introduced $\alpha_i \equiv N_a p_i \cdot q$ and $\beta_j \equiv N'_b p_j \cdot q'$. Exploiting the residual $\mathrm{SL}(2,\mathbb{R})$ gauge invariance of the worldsheet to fix three out of four $w_i$ values, we end up with six channels of the amplitude, labeled in terms of the Mandelstam variables $s,t,u$ defined the usual way:\footnote{The definition is given in the description of kinematics, in Eq.~(\ref{mandelstam}). At this place we do not need the definition of Mandelstam variables, we just want to emphasize that, depending on the permutations of the insertion points on the worldsheet, we get six different channels; for this reason we only give the definitions of $s,t,u$ later in the text, when we have defined all the momenta.}
\be
\mathcal{A} = \mathcal{A}_{st} +\mathcal{A}_{tu}+\mathcal{A}_{us}+\mathcal{A}_{ts}+\mathcal{A}_{ut}+\mathcal{A}_{su}.\label{astu}
\ee
It is enough to state in full the expression for one channel; the others are then obtained by simple permutations of the momenta.\footnote{We write the amplitude without Chan-Patton factors so it reduces to the sum of the six cyclic orderings; equivalently, one can imagine Chan-Patton factors of an Abelian $U(1)$ group which reduce to just an overall multiplication of the sum. It would be interesting to consider nontrivial Chan-Patton factors as additional group structure could well further reduce chaos and divide the S-matrix into symmetry sectors. But at the present level of understanding this would be a superfluous complication.}  We give the expression for the $st$ channel:
\bea
\mathcal{A}_{st} &=& \mathcal{A}\vert_{w'_1 = -\infty, w_2 = 0, w_1 = w, w'_2 = 1} = \nonumber\\
&=&\int_0^{1} dw w^{p_1 \cdot p_2} (1-w)^{p_1 \cdot p'_2} \prod_{a = 1}^{J} Z_a^{212'}(\alpha,p,\lambda; w) \prod_{b = 1}^{J'} Z_b^{212'}(\beta,p,\lambda'; w),\label{astmain}
\eea
where the integrals $Z_a^{ijk}$ are defined in \cite{Hashimoto:2022bll} as
\bea
&&Z_a^{ijk}(\alpha,p,\lambda;w)\equiv Z_a(\alpha_i,\alpha_j,\alpha_k,p,\lambda;w) \equiv \nonumber\\
&\equiv&\int_{-\infty}^{\infty} dz_a \vert z_a\vert^{-\alpha_i} \vert z_a - w\vert^{-\alpha_j} \vert z_a - 1\vert^{-\alpha_k} \bigg( \frac{-p_i \cdot \lambda}{-z_a} + \frac{-p_j \cdot \lambda}{w - z_a} + \frac{-p_k \cdot \lambda}{1 - z_a} \bigg).\label{zijk}
\eea
The other channels are now related to the $st$ channel in the following way:
\begin{eqnarray}
  \mathcal{A}_{tu} &=& \mathcal{A}\vert_{w_2 = -\infty, w'_2 = 0, w_1 = w, w'_1 = 1} = \mathcal{A}_{st}\vert_{2 \rightarrow 2', 2' \rightarrow 1'},\label{atu}\\
  \mathcal{A}_{us} &=& \mathcal{A}\vert_{w'_2 = -\infty, w'_1 = 0, w_1 = w, w_2 = 1} = \mathcal{A}_{st}\vert_{2 \rightarrow 1', 2' \rightarrow 2},\label{aus}\\
  \mathcal{A}_{ts} &=& \mathcal{A}_{st}\vert_{2 \longleftrightarrow 2'},\label{ats}\\
  \mathcal{A}_{ut} &=& \mathcal{A}_{tu}\vert_{2' \longleftrightarrow 1'}\label{aut},\\
  \mathcal{A}_{su} &=& \mathcal{A}_{us}\vert_{1' \longleftrightarrow 2}.\label{asu}
\end{eqnarray}
The on-shell condition needed to create $\mathrm{HES}$ and $\mathrm{HES'}$ is that after the first $j \leq J$ photons have been absorbed, the mass of the intermediate state is $M_j = 2 (\sum_{a \leq j} N_a  -1)$. Having in mind that the total momentum of this intermediate state is $p_2 - \sum_{a \leq j} q$, and using the on-shell condition for tachyons $-p_2^2 = -2$, we obtain:
\be
\alpha_2 \rightarrow  N_a, \beta_2' \rightarrow N'_b 
\label{onshell}
\ee
where the second condition comes from $\mathrm{HES'}$. The $Z_a^{ijk}$ integrals can be written in terms of regularized hypergeometric functions ${}_2\tilde{F}_1(a,b;c;z)\equiv {}_2F_1(a,b;c;z)/\Gamma(c)$. The coefficients of their Taylor expansions are then expressed in terms of gamma functions. This calculation is thoroughly described in \cite{Hashimoto:2022bll}, the result for the $st$ channel being
\be
\mathcal{A}_{st} = \sum_{i_a \in \{2,2'\}} \sum_{j_b \in \{2,2'\}} \sum_{k_a = 1}^{N_a} \sum_{l_b = 1}^{N'_b} \bigg ( \prod_{a}^{J} (p_{i_a} \cdot \lambda) 
 c_{k_a}^{(i_a)} \bigg) \bigg ( \prod_{b}^{J'} (p_{j_b} \cdot \lambda') 
 d_{l_b}^{(j_b)} \bigg) B\left(- 1 - \frac{s}{2} + k, -1 -\frac{t}{2} + l\right),\label{ast}
\ee
where $l\equiv\sum_b^{J'}l_b$, $k\equiv\sum_a^Jk_a$, the coefficients $c$ and $d$ are defined as:
\begin{eqnarray}
  &&c_k^{(2)} = c_k(\alpha_2+1,\alpha_1 + 1, \alpha'_2),\label{coeffc2}\\
  &&c_k^{(2')} = -c_{k-1}(\alpha_2,\alpha_1, \alpha'_2+1),\label{coeffc2p}\\
  &&d_l^{(2)} = -c_{l-1}(\beta'_2,\beta_1, \beta_2+1),\label{coeffd2}\\
  &&d_l^{(2')} = c_l(\beta'_2+1,\beta_1+1, \beta_2),\label{coeffd2p}
\end{eqnarray}
using the function:
\be
c_k(\alpha_2,\alpha_1,\alpha'_2)  = (-1)^{\alpha_2 + k - 1}\frac{\pi}{\sin\left(\pi\alpha_2\right)}\frac{\Gamma(\alpha_2 -k +\alpha_1 - 1) \Gamma(\alpha'_2 + k)}{\Gamma(\alpha_1) \Gamma(\alpha'_2) \Gamma(\alpha_2 - k) \Gamma(k+1)},
\label{ckdef}
\ee
and finally the beta function is defined as usual: $B(x,y)\equiv\Gamma(x)\Gamma(y)/\Gamma(x+y)$.

Because of the condition (\ref{onshell}), the coefficients $c$ and $d$ from Eqs.~(\ref{coeffc2}-\ref{coeffd2p}) diverge as we can see from the factor $\pi/\sin(\pi\alpha_2)$ in Eq.~(\ref{ckdef}). This divergence is just the expected behavior of the amplitude when one of the internal momenta in the diagram goes on-shell. In order to extract the amplitude of the $\mathrm{HES}+\mathrm{t}\longrightarrow\mathrm{HES'}+\mathrm{t'}$ process we thus have to regularize the amplitude. To achieve this we simply omit the divergent factor $\pi/\sin(\pi\alpha_2)$ in the definition (\ref{ckdef}), resulting in a finite expression for the amplitude (except of course for special choices of kinematic variables).\footnote{This procedure is essentially the same as the one in quantum field theory where one would remove the divergence by multiplying the amplitude by the inverse propagator of the internal on-shell particle.}

For a generic partition $\sum_{a = 1}^{J} N_a = N$ the number of operations needed to compute the amplitude (\ref{ast}) increases rapidly with the partition lengths $J$ and $J'$ due to the increase of the number of sums over $k_a$ and $l_b$. In addition, the number of states with fixed level $N$ grows exponentially in $\sqrt{N}$. These two effects make the calculation of the S-matrix in the whole subspace of fixed $N$ and fixed kinematic variables computationally demanding, heavily limiting the maximum value of $N$ we can work with and requiring the use of a cluster for larger values of $N$.

%On the other hand, in a generic quantum field theory, the number of states does not increase with energy so rapidly, providing only a small (of order $O(1)$) number of channels through which chaos can develop. Hence one would expect chaos to emerge only at strong coupling in most field theories. In other words, even though the large number of $\mathrm{HES}$ states makes the problem computationally demanding, it is also the reason why even strong chaos may be probed analytically (at least in principle) in the $\mathrm{HES}$ scattering.

\subsubsection{Kinematics}\label{sechesopenkin}

A look at the expression (\ref{ast}) for $\mathcal{A}_{st}$ tells that the kinematics is highly non-unique: there are many momenta and polarizations involved and we have many parameters to choose. We have made no attempt to consider their influence in full detail. We have varied the momenta and the scattering angles (one at a time) over some representative intervals; with some hindsight, we can say that for tachyon scattering only the magnitude of the incoming momentum is crucial. As noted above, $p_1$ and $p'_1$ are the momenta of the on-shell tachyons while $\mathrm{HES}$ is created from the tachyon with momentum $p_2$ and a set of $J$ photons with momenta $\{-N_a q\}$ whose polarizations are all equal to $\lambda$ (similarly for the $\mathrm{HES'}$ we have $p'_2, J', \{N'_b\},q'$ and $\lambda'$). The on-shell conditions and the momentum conservation equation now read:
\bea
&&p_1^2=p_2^2=\left(p'_1\right)^2=\left(p'_2\right)^2=2,~~q^2=q'^2=0\\
&&p_1 + (p_2 -N q) + p'_1 + (p'_2 - N q')=0.\label{momentumcons}
\eea
The total mass of $\mathrm{HES}$ (and $\mathrm{HES'}$) is given by the total occupation number $M = 2(N-1) = 2(N'-1)$, while $J$ and $J'$ represent the total spin of $\mathrm{HES}$ and $\mathrm{HES'}$ respectively because the photons have identical polarizations. We choose to work in the center-of-mass frame and, as we already mentioned, for simplicity we take the polarizations and photon momenta to satisfy $\lambda = -\lambda'$, $\lambda\cdot\lambda = 0$ and $q' \propto q$. In order to satisfy these conditions we parametrize momenta and polarizations in the following way:
\begin{eqnarray}
  q &=& \frac{1}{\sqrt{2(N-1) + p^2} - p \cos \theta} (-1, 0, 0, 1)^{T}\nonumber\\
  q' &=& \frac{1}{\sqrt{2(N-1) + p^2} - p \cos \theta'} (1, 0, 0, -1)^{T}\nonumber\\
  p_1 &=& \left(\sqrt{p^2-2}, p \sin \theta, 0, p \cos \theta\right)^{T}\nonumber\\
  p'_1 &=& -\left(\sqrt{p^2-2}, p \sin \theta' \cos \phi', p\sin \theta' \sin \phi', p \cos \theta'\right)^{T}\nonumber\\ 
  %p_2 &=& N q - p_1 = \left(\sqrt{s}, 0, 0, 0\right)^{T},  p'_2 = N'q' -p'_1 -\left(\sqrt{s}, 0, 0, 0\right)^{T}\nonumber\\ 
  p_2 &=& \left(\sqrt{2N-2+p^2}-\frac{N}{\sqrt{2N-2+p^2}-p\cos\theta},-p\sin\theta,0,-p\cos\theta+\frac{N}{\sqrt{2N-2+p^2}-p\cos\theta}\right)^T\nonumber\\
  p'_2 &=& \Bigg(-\sqrt{2N'-2+p^2}+\frac{N'}{\sqrt{2N'-2+p^2}-p\cos\theta'},p\sin\theta'\cos\phi',p\sin\theta'\sin\phi',\nonumber\\
  &&p\cos\theta'-\frac{N}{\sqrt{2N-2+p^2}-p\cos\theta'}\Bigg)^T\nonumber\\
  \lambda &=& \frac{1}{\sqrt{2}} (0, 1, i, 0)^T.\label{kinemeqs}
\end{eqnarray}
Defining the Mandelstam variables in terms of momenta:
\bea
%  s &=& -\left(p_1 + p_2 - N q\right)^2 = \left(\sqrt{2(N-1) + p^2-2} +\sqrt{p^2-2}\right)^2\\
%  t &=& -\left(p_1 +p'_2 - N' q'\right)^2 = \left(\sqrt{2(N-1) - p^2-2} +\sqrt{p^2-2}\right)^2-\nonumber\\
%  &&- 2p^2\left(1+\cos \theta \cos\theta' + \sin \theta \sin \theta' \cos \phi'\right)\\
%  u &=& -\left(p_1 + p'_1\right)^2 = -2p^2\left(1 - \cos \theta \cos\theta' - \sin \theta \sin \theta' \cos \phi'\right).
%s &=& -\left(p_1 + p_2 - N q\right)^2\nonumber\\
%t &=& -\left(p_1 +p'_2 - N' q'\right)^2\nonumber\\
%u &=& -\left(p_1 + p'_1\right)^2.\label{mandelstam}
s=-\left(p_1 + p_2 - N q\right)^2,~t=-\left(p_1+p'_2-N'q'\right)^2,~u=-\left(p_1+p'_1\right)^2,\label{mandelstam}
\eea
%(e.g. $\alpha_2 = N_a p_2 \cdot q \rightarrow N_a$)
we obtain $p_2 \cdot q = p'_2 \cdot q' = 1$ as required by the on-shell conditions (\ref{onshell}). Furthermore, the only imaginary contribution to $\mathcal{A}_{st}$ comes from the $(p\cdot \lambda)$ factors which we calculate to be $p'_2 \cdot \lambda' = -\frac{p\sin\theta'}{\sqrt{2}} e^{i \phi'}$. Hence for $\phi' = 0$ the amplitudes calculated from Eq.~(\ref{ast}) will be real. For the collinear kinematics, that is $\theta, \theta' \in \{0,\pi\}$, the amplitude vanishes as can be seen from the fact that $\lambda$ is orthogonal to all momenta.

From the above, the amplitude is fully characterized by the module of the momentum $p$, scattering angles $\theta, \theta'$ and $\phi'$, and by the partitions of the levels ${N_a}$ and ${N'_b}$. The amplitudes then define the elements of the S-matrix for fixed kinematics ($p, \theta, \theta', \phi'$) in the basis of different partitions of the level $N$.

\subsection{Open HES-photon amplitudes}\label{sechesspin}

It will turn out crucial to check also the dynamics of a spinful (non-scalar) state scattering on the HES -- some novel aspects of transient chaos will only show up for initial states with spin. We thus consider the HES-photon scattering process (\ref{processgamma}).\footnote{As we have mentioned, the HES-graviton process is obtained by applying the KLT relations to the HES-photon process.} The DDF construction of HES states proceeds exactly the same way as before, with the same assumptions and conditions on the momenta $q$, $q'$ and polarizations $\lambda$, $\lambda'$ of DDF photons and the momenta $p_2$, $p_2'$ of DDF tachyons. The sole difference lies in the states $1,1'$ which now describe photons with vertex operators $V_\gamma$, momenta $p_1$, $p'_1$ satisfying $p_1^2=\left(p'_1\right)^2=0$ and polarizations $\xi_1\equiv\xi,\xi_{1'}\equiv\xi'$ satisfying the gauge invariance condition $\xi\cdot p_1=\xi'\cdot p'_1=0$. The amplitude is now
\bea
\mathcal{A}^\gamma &=& \frac{1}{\mathrm{Vol}\left(\mathrm{SL}\left(2,\mathbb{R}\right)\right)}\int\mathcal{D}X e^{-S_\mathrm{P}}\int \prod_i dw_i V_t(w_i,p_i)\int \prod_K dw_K V_\gamma(w_K,p_K) \times\nonumber\\
&\times&\prod_{a=1}^J dz_a V_p(z_a,-N_a q,\lambda) \prod_{b=1}^{J'} dz'_b V_p(z'_b,-N'_b q,\lambda),\label{ampphoton}
\eea
where now $i,j\in\lbrace 2,2'\rbrace$ are DDF tachyons, $K,L\in\lbrace 1,1'\rbrace$ are the physical photons, and the rest of the notation is the same as in Eq.~(\ref{amptachyon}). Writing out the insertions of the vertex operators, we can write the amlitude as
\be 
\mathcal{A}^\gamma = \frac{1}{\mathrm{Vol}\left(\mathrm{SL}\left(2,\mathbb{R}\right)\right)}\int_{-\infty}^{\infty} \prod_{i} dw_i \prod_{K} dw_K \int_{-\infty}^{\infty}\prod_{a=1}^J dz_a \int_{-\infty}^{\infty}\prod_{b=1}^{J'} dz'_b~\frak{a}_\mathrm{I}\frak{a}_\mathrm{II}\frak{a}_\mathrm{III},\label{openampphoton}
\ee
where the factors $\frak{a}_\mathrm{I,II,III}$ denote the tachyon-tachyon, tachyon-photon and photon-photon terms respectively:
\bea
%tach-tach
\frak{a}_\mathrm{I}&=&\prod_{i < j} \vert w_i-w_j\vert^{p_i \cdot p_j}\nonumber\\
%tach-phot
\frak{a}_\mathrm{II}&=&\prod_{i,K} \vert w_i-w_K\vert^{p_i \cdot p_K} \sum_{i}\frac{-p_i\cdot\xi_K}{w_i-w_K}
\prod_{i,a} \vert z_a-w_i\vert^{-\alpha_i}\sum_{i}\frac{-p_i\cdot\lambda}{w_i-z_a}
\prod_{j,b} \vert z'_b-w_j\vert^{-\beta_j}\sum_{j}\frac{-p_j\cdot\lambda'}{w_j-z'_b}\nonumber\\ 
%phot-phot
\frak{a}_\mathrm{III}&=&\prod_{K,a}\vert z_a-w_K\vert^{-\alpha_K}\prod_{L,b} \vert z'_b-w_L\vert^{-\beta_L}\prod_{K<L}\vert w_K-w_L\vert^{p_K\cdot p_L}\times\exp\Bigg[\sum_{K<L}\frac{\xi_K\cdot\xi_L}{2w_{KL}^2}+\nonumber\\
&\times&\sum_{K,a}\left(\frac{-p_a\cdot\xi_K+p_K\cdot\lambda}{w_K -z_a}+\frac{\xi_K\cdot\lambda}{2\vert w_K-z_a\vert^2}\right)+\sum_{L,b}\left(\frac{-p_b\cdot\xi_L+p_L\cdot\lambda'}{w_L-z'_b}+\frac{\xi_L\cdot\lambda'}{2\vert w_L-z_b\vert^2}\right)\Bigg].\nonumber\\ \label{openamp3}
\eea
%\frak{a}_\mathrm{III}&=&\prod_{K,a} \vert z_a-w_K\vert^{-\alpha_K}\sum_{K,a}\left(\frac{-p_a\cdot\xi_K+p_K\cdot\lambda}{w_K -z_a}+\frac{\xi_K\cdot\lambda}{2\vert w_K-z_a\vert^2}\right)\times\nonumber\\
%&\times&\prod_{L,b} \vert z'_b-w_L\vert^{-\beta_L} \sum_{L,b}\left(\frac{-p_b\cdot\xi_L+p_L\cdot\lambda'}{w_L-z'_b}+\frac{\xi_L\cdot\lambda'}{2\vert w_L-z_b\vert^2}\right)\times\nonumber\\
%&\times&\prod_{K<L}\vert w_K-w_L\vert^{p_K\cdot p_L}\sum_{K<L}\frac{\xi_K\cdot\xi_L}{2w_{KL}^2}.\label{openamp3}
Here $\alpha_i$, $\alpha_K$, $\beta_j$, $\beta_L$ have the same meaning as before (taking into account the definition of $i,j,K,L$ above), and it is understood as usual that we only take the terms in the expansion of the exponent which are linear in every single polarization. The integrals in Eq.~(\ref{openampphoton}) can again be expressed in terms of the same functions $Z_a^{ijk}$ from Eq.~(\ref{zijk}), and different permutations of the insertion points on the worldsheet again produce the sum of contributions for various permutations as in Eqs.~(\ref{astu},\ref{atu}-\ref{asu}). But now the photon-photon interaction terms, encapsulated in $\frak{a}_\mathrm{III}$ from Eq.~(\ref{openamp3}) cannot in general be avoided: we cannot in general impose the additional conditions on the physical photons $1,1'$ akin to those for DDF photons as we would not have enough equations to satisfy the momentum conservation. Therefore, many additional terms will appear in the expression for each channel.
%The general expression is:
%\bea
%\mathcal{A}&=&\prod_i\int dw_i\prod_K\int dw_K\prod_a\int dz_a\prod_b\int dz'_b\times\nonumber\\
%&\times&\prod_{i<j}\vert w_i-w_j\vert^{p_i\cdot p_j}\prod_{K<L}\vert w_K-w_L\vert^{p_K\cdot p_L}\prod_{i,K}\vert w_i-w_K\vert^{p_i\cdot p_K}\times\nonumber\\
%&\times&\prod_{i,a}\vert z_a-w_i\vert^{p_i\cdot q}\prod_{i,b}\vert z'_b-w_i\vert^{p_i\cdot q'}\prod_{K,a}\vert z_a-w_K\vert^{p_K\cdot q}\prod_{K,b}\vert z'_b-w_K\vert^{p_K\cdot q'}\times\nonumber\\
%&\times&\prod_K\sum_i\frac{-p_i\cdot\xi_K}{w_i-w_K}\prod_a\sum_i\frac{-p_i\cdot\lambda}{w_i-z_a}\prod_b\sum_i\frac{-p_i\cdot\lambda}{w_i-z'_b}\times\nonumber\\
%&\times&\exp\Bigg[\sum_{K\neq L}\frac{-p_K\cdot\xi_L}{w_K-w_L}+\nonumber\\
%&+&\Bigg]
%\eea
We again give the expression for $\mathcal{A}_{st}$, understanding that the others are obtained by permutations of the indices $1,2,1',2'$:
\bea
\mathcal{A}^\gamma_{st} &=& \mathcal{A}\vert_{w'_1 = -\infty, w_2 = 0, w_1 = w, w'_2 = 1} = \int_0^{1} dw w^{p_1 \cdot p_2} (1-w)^{p_1 \cdot p'_2}\times\nonumber\\
&\times&\Bigg[\left(\frac{p_2\cdot\xi_1}{w}+\frac{p'_2\cdot\xi_1}{1-w}\right)\prod_{a = 1}^{J} Z_a^{212'}(\alpha,p,\lambda; w) \prod_{b = 1}^{J'} Z_b^{212'}(\beta,p,\lambda'; w)+\nonumber\\
&+&\left(q\cdot\xi_1-p_1\cdot\lambda\right)\prod_{a = 1}^{J} Z_a^{212'}(\alpha,p,\lambda; w;-1) \prod_{b = 1}^{J'} Z_b^{212'}(\beta,p,\lambda'; w)+\nonumber\\
&+&\left(q'\cdot\xi_1-p_1\cdot\lambda'\right)\prod_{a = 1}^{J} Z_a^{212'}(\alpha,p,\lambda; w) \prod_{b = 1}^{J'} Z_b^{212'}(\beta,p,\lambda'; w;-1)+\nonumber\\
&+&\left(\xi_1\cdot\lambda\right)\prod_{a = 1}^{J} Z_a^{212'}(\alpha,p,\lambda; w;-2) \prod_{b = 1}^{J'} Z_b^{212'}(\beta,p,\lambda'; w)+\nonumber\\
&+&\left(\xi_1\cdot\lambda'\right)\prod_{a = 1}^{J} Z_a^{212'}(\alpha,p,\lambda; w) \prod_{b = 1}^{J'} Z_b^{212'}(\beta,p,\lambda'; w;-2)\Bigg].\label{astmainphoton}
\eea
Here we introduce the notation $Z_a^{212'}(\alpha,p,\lambda;w;-\sigma)$ with $\sigma$ a positive integer. This is defined as
\be 
%Z_a^{iKj}(\alpha,p,\lambda; w;-\sigma)\equiv Z_a^{iKj}(\alpha_i-\sigma,\alpha_K,\alpha_j,p,\lambda; w)+Z_a^{iKj}(\alpha_i,\alpha_K,\alpha_j-\sigma,p,\lambda; w).\label{zijkphoton}
Z_a^{iKj}(\alpha,p,\lambda; w;-\sigma)\equiv Z_a(\alpha_i-\sigma,\alpha_K,\alpha_j,p,\lambda; w)+Z_a(\alpha_i,\alpha_K,\alpha_j-\sigma,p,\lambda; w).\label{zijkphoton}
\ee
In other words, the argument $-\sigma$ means we sum over the values of the original $Z^{iKj}_a$ as defined in Eq.~(\ref{zijk}) with the arguments $\alpha_i,\alpha_j$ being reduced by $\sigma$ one at a time. Notice that we do not include the function with $\alpha_K-\sigma$ in the sum in (\ref{zijkphoton}), i.e. we only reduce the $\alpha$'s corresponding to DDF tachyons, not physical photons. The outcome is that the photon amplitude $\mathcal{A}_{st}^\gamma$ consists of the same building blocks as the tachyon amplitude $\mathcal{A}_{st}$ given in Eqs.~(\ref{astmain},\ref{ast}) but with different coefficients and arguments of the gamma and beta functions. Writing out in full the expression (\ref{astmainphoton}) we obtain:
\be
\mathcal{A}^\gamma_{st}=\sum_{i_a \in \{2,2'\}} \sum_{K_b\in \{2,2'\}} \sum_{k_a = 1}^{N_a} \sum_{l_b = 1}^{N'_b}\left(\frak{b}_0+\frak{b}_{-1}+\frak{b}'_{-1}+\frak{b}_{-2}+\frak{b}'_{-2}\right).\label{astphoton}
\ee
Note that the terms $\frak{b}_{0,-1,-2}$ and $\frak{b}'_{-1,-2}$ are not related in any simple way to $\frak{a}_\mathrm{I,II,III}$ from Eqs.~(\ref{openamp3}). They are defined as:
\bea
\frak{b}_0&=&\bigg(\prod_{a}^{J}(p_{i_a}\cdot\lambda)c_{k_a}^{(i_a)}\bigg)\bigg (\prod_{b}^{J'}(p_{K_b}\cdot\lambda')d_{l_b}^{(K_b)}\bigg)\Bigg[\left(p_{i_a}\cdot\xi_{K_b}\right)B\left(-2-\frac{s}{2}+k,-1-\frac{t}{2}+l\right)+\nonumber\\
&&+\left(p_{i_a}\cdot\xi_{K_b}\right)B\left(-1-\frac{s}{2}+k,-2-\frac{t}{2}+l\right)\Bigg]\label{astphoton0}\\
\frak{b}_{-1}&=&\left(q\cdot\xi_{K_b}-p_{i_a}\cdot\lambda\right)\Bigg[\bigg(\prod_{a}^{J}(p_{i_a}\cdot\lambda)\bar{c}_{k_a}^{('i_a)}\bigg)\bigg (\prod_{b}^{J'}(p_{K_b}\cdot\lambda')d_{l_b}^{(K_b)}\bigg)B\left(-1-\frac{s}{2}+k,-2-\frac{t}{2}+l\right)+\nonumber\\
&&+\bigg(\prod_{a}^{J}(p_{i_a}\cdot\lambda)\bar{c}_{k_a}^{('''i_a)}\bigg)\bigg (\prod_{b}^{J'}(p_{K_b}\cdot\lambda')d_{l_b}^{(K_b)}\bigg)B\left(-2-\frac{s}{2}+k,-1-\frac{t}{2}+l\right)\Bigg]\label{astphoton1}\\
\frak{b}'_{-1}&=&\left(q'\cdot\xi_{K_b}-p_{i_a}\cdot\lambda'\right)\bigg(\prod_{a}^{J}(p_{i_a}\cdot\lambda)c_{k_a}^{(i_a)}\bigg)\bigg(\prod_{b}^{J'}(p_{K_b}\cdot\lambda')\bar{d}_{l_b}^{(''K_b)}\bigg)B\left(-1-\frac{s}{2}+k,-1-\frac{t}{2}+l\right)\nonumber\\
&&\label{astphoton1p}\\
\frak{b}_{-2}&=&\left(\xi_{K_b}\cdot\lambda\right)\Bigg[\bigg(\prod_{a}^{J}(p_{i_a}\cdot\lambda)\bar{\bar{c}}_{k_a}^{('i_a)}\bigg)\bigg (\prod_{b}^{J'}(p_{K_b}\cdot\lambda')d_{l_b}^{(K_b)}\bigg)B\left(-1-\frac{s}{2}+k,-3-\frac{t}{2}+l\right)+\nonumber\\
&&+\bigg(\prod_{a}^{J}(p_{i_a}\cdot\lambda)\bar{\bar{c}}_{k_a}^{('''i_a)}\bigg)\bigg (\prod_{b}^{J'}(p_{K_b}\cdot\lambda')d_{l_b}^{(K_b)}\bigg)B\left(-3-\frac{s}{2}+k,-1-\frac{t}{2}+l\right)\Bigg]\label{astphoton2}\\
\frak{b}'_{-2}&=&\left(\xi_{K_b}\cdot\lambda'\right)\bigg(\prod_{a}^{J}(p_{i_a}\cdot\lambda)c_{k_a}^{(i_a)}\bigg)\bigg(\prod_{b}^{J'}(p_{K_b}\cdot\lambda')\bar{\bar{d}}_{l_b}^{(''K_b)}\bigg)B\left(-2-\frac{s}{2}+k,-2-\frac{t}{2}+l\right).\label{astphoton2p}
\eea
The coefficients $c_k^{(i)}$, $d_l^{(K)}$ and the function $c_k(\alpha_i,\alpha_K,\alpha_j)$ have the same meaning as in Eqs.~(\ref{coeffc2}-\ref{ckdef}). The newly introduced coefficients are defined as:
\begin{eqnarray}
&&\bar{c}_k^{('2)} = c_k(\alpha_2,\alpha_1 + 1, \alpha'_2),~\bar{c}_k^{('''2)} = c_k(\alpha_2+1,\alpha_1 + 1, \alpha'_2-1),\label{coeffbarc2}\\
&&\bar{c}_k^{('2')} = -c_{k-1}(\alpha_2-1,\alpha_1, \alpha'_2+1),~\bar{c}_k^{('''2')} = -c_{k-1}(\alpha_2,\alpha_1, \alpha'_2),\label{coeffbarc2p}\\
&&\bar{d}_l^{(''2)} = -c_{l-1}(\beta'_2,\beta_1-1,\beta_2+1),\label{coeffbard2}\\
&&\bar{d}_l^{(''2')} = c_l(\beta'_2+1,\beta_1, \beta_2)\label{coeffbard2p}
\end{eqnarray}
and analogously for the two-bar coefficients:
\begin{eqnarray}
&&\bar{\bar{c}}_k^{('2)} = c_k(\alpha_2-1,\alpha_1 + 1, \alpha'_2),~\bar{\bar{c}}_k^{('''2)} = c_k(\alpha_2+1,\alpha_1 + 1, \alpha'_2-2),\label{coeffbarbarc2}\\
&&\bar{\bar{c}}_k^{('2')} = -c_{k-1}(\alpha_2-2,\alpha_1, \alpha'_2+1),~\bar{\bar{c}}_k^{('''2')} = -c_{k-1}(\alpha_2,\alpha_1, \alpha'_2-1),\label{coeffbarbarc2p}\\
&&\bar{\bar{d}}_l^{(''2)} = -c_{l-1}(\beta'_2,\beta_1-2,\beta_2+1),\label{coeffbarbard2}\\
&&\bar{\bar{d}}_l^{(''2')} = c_l(\beta'_2+1,\beta_1-1, \beta_2).\label{coeffbarbard2p}
\end{eqnarray}
In other words, the number of bars (one or two) corresponds to the value of $\sigma$ in the function $Z_a^{iKj}(\alpha,p,\lambda; w;-\sigma)$ in Eq.~(\ref{zijkphoton}), and the additional superscript $'$, $''$ or $'''$ means we should subtract $\sigma$ from the first, second or third argument of the function $c_k(\alpha_i,\alpha_K,\alpha_j)$ from Eq.~(\ref{ckdef}).

Finally, we should define the kinematics. We minimally modify the setup for the HES-tachyon process in the subsubsection \ref{sechesopenkin}. Energy and momentum conservation now read
\bea
&&p_2^2=\left(p'_2\right)^2=2,~~p_1^2=\left(p'_1\right)^2=q^2=q'^2=0\\
&&p_1 + (p_2 -N q) + p'_1 + (p'_2 - N q')=0.\label{momentumconsphoton}
\eea
The momenta and polarizations of the physical photons are now given by
\begin{eqnarray}
  p_1 &=& \left(p, p \sin \theta, 0, p \cos \theta\right)^{T}\nonumber\\
  p'_1 &=& -\left(p, p \sin \theta' \cos \phi', p\sin \theta' \sin \phi', p \cos \theta'\right)^{T}\\
  \xi_1 &\equiv &\xi = (0,\cos\theta,0,-\sin\theta)^T \nonumber\\
  \xi'_1 &\equiv &\xi'=(0,-\cos\theta',0,\sin\theta'\cos\phi'),\label{kinemeqsphoton}
\end{eqnarray}
whereas all the other momenta are the same as for the HES-tachyon scattering, as given in Eq.~(\ref{kinemeqsphoton}). The simplifications which eliminate the interaction terms among the DDF photons are still in place. For the physical photons they are absent (indeed, it seems the number of free parameters is insufficient to require $\xi'\propto\xi$ after implementing the conservation laws) therefore we indeed have photon-photon terms in the amplitudes as we found in Eqs.~(\ref{openamp3}-\ref{astphoton2p}).
 
%We can again lose the terms with products of polarizations $\xi_K\cdot\xi_L$, $\xi_K\cdot\lambda$ and $\xi_K\cdot\lambda'$ if we choose the polarizations of physical photons $\xi_1$ and $\xi_2$ to be equal to $\lambda$ and $\lambda'$: then we have $\lambda^2=\lambda'^2=\xi_K^2=\lambda\cdot\lambda'=\lambda\cdot\xi_K=\lambda'\cdot\xi_K=0$ and the term $\frak{a}_\mathrm{III}$ becomes
%\bea
%\frak{a}_\mathrm{III}&=&\prod_{K<L}\vert w_K-w_L\vert^{p_K\cdot p_L}\times\nonumber\\
%&\times&\prod_{K,a}\vert z_a-w_K\vert^{-\alpha_K}\sum_{K,a}\left(\frac{-p_a\cdot\xi_K+p_K\cdot\lambda}{w_K -z_a}\right)\prod_{L,b} \vert z'_b-w_L\vert^{-\beta_L} \sum_{L,b}\left(\frac{-p_b\cdot\xi_L+p_L\cdot\lambda'}{w_L-z'_b}\right).~~~~~~~~~~\label{openamp30}
%\eea

\subsection{Closed string amplitudes}\label{sechesclosed}

In the previous section we have described the HES-tachyon scattering for open (bosonic) HES. This process already shows interesting physics as we will see, however to make closer contact with black holes or with the results on classical string scattering in the literature \cite{Frolov1999,BasuAdSS,BasuRN}, we should also consider closed string amplitudes. This can be achieved in a similar way as for the open case, making use of the DDF operators. But at tree level we can circumvent this calculation by employing the celebrated KLT relations \cite{KLT,Sondergaard:2011ivKLT,Stieberger:2014hbaKLT}. To remind, the idea behind the KLT relations is that a closed string amplitude can be constructed from two open string amplitudes coupled by a momentum-dependent kernel.\footnote{The connection between open and closed tree-level amplitudes can be seen from the fact that the closed string propagator can be constructed by joining the two open string propagators.} Schematically, an $M$-point closed string amplitude $\mathcal{A}_\mathrm{closed}^{M}$ is given by:
\be
\mathcal{A}_\mathrm{closed}^{M} \propto \sum_{P,P'} \mathcal{A}(P)_\mathrm{open}^{M} \mathcal{\bar A}(P')_\mathrm{open}^{M} e^{i F(P,P')}
\ee
where $P$ and $P'$ denote the permutations of the $M$ external legs, $\mathcal{A}(P)_\mathrm{open}^{M}$ is the ordered $M$-point open string amplitude and $F$ is a phase determined by the kinematics. %In the field theory limit ($\alpha' \rightarrow 0$), for $M = 3$, one recovers that the three-point graviton vertex is equal to the product of the two color-stripped Yang-Mills three-point vertices, an example of the double copy relation between field theories \cite{PhysRevLett.105.061602DblCopy}.

In the case of interest for us, that is for the four-leg ($2\to 2$) scattering, the KLT relation becomes particularly simple:
\be
\zeta_{\mu_1\ldots\mu_4,\nu_1,\ldots\nu_4} \mathcal{A}_\mathrm{closed}^{\mu_1\ldots\mu_4,\nu_1\ldots\nu_4} = -\pi \sin(\pi p_2 \cdot p'_1) \xi_{\mu_1\ldots\mu_4} \mathcal{A}_\mathrm{open}^{\mu_1\ldots\mu_4} (s,t)\xi'_{\mu_1\ldots\mu_4} \mathcal{A'}_\mathrm{open}^{\nu_1\ldots\nu_4}(t,u).
\label{klt4}
\ee
This enables us to construct the closed string S-matrix as a direct product of the $st$ and $tu$ contributions to the open string S-matrix. If we consider fixed kinematics and vary the partitions, the direct product structure changes the dimensions of the S-matrix from $p(N) \times p(N)$ for open to $p(N)^2 \times p(N)^2$ for closed strings, which is a big enhancement in size having in mind the fast growth of the number of partitions $p(N)$ with $N$. In this way we obtain the closed HES-tachyon and the closed HES-graviton processes of Eqs.~(\ref{processc},\ref{processgrav}) from the open HES processes of Eqs.~(\ref{process},\ref{processgamma}) respectively.

For our main interest -- inspecting the eigenphase statistics and chaos of the S-matrix -- the KLT method may be potentially risky as it yields the closed string amplitudes in the uncorrelated basis, i.e. the closed string states are expressed in terms of direct products of open-string states $\vert\vec{n}\rangle\otimes\vert\vec{\tilde{m}}\rangle$ projected to the subspace satisfying $N=\tilde{N}$, i.e. $\sum_an_a=\sum_b\tilde{m}_b$. In this basis the symmetry between the left- and right-moving modes is not manifest so even and odd states under this symmetry are dumped together. It is known that expressing the Hamiltonian/S-matrix in a basis which is not adapted to the symmetries of the system can invalidate the results in the sense that Wigner-Dyson statistics does not show up even if the system is in fact RMT chaotic. We have thus checked our results for small occupation numbers $N\leq 6$ against numerically computed S-matrices in the even and odd sector separately (obtained by performing the worldsheet integrals via a grid method). We find no discrepancy between the resulting eigenphase statistics, however for larger $N$ we were unable to perform this test as the numerical integration becomes unfeasible. Nevertheless, we consistently find equal or stronger chaos for closed than for open HES, meaning that that there is no artifical reduction of chaos because the spectrum unfolding is not performed.\footnote{According to \cite{Bianchi:2022mhs}, testing the spacing ratios instead of spacings themselves is also helpful in such situations the ratios $r$ and $R$ are less sensitive to folding/unfolding of the spectrum, and in our case again they give similar results as the Wigner-Dyson fits themselves.}

\section{The structure of the S-matrix}\label{secchaos}

Our basic object is the S-matrix for the processes (\ref{process}-\ref{processgamma}) involving two HES states and the two tachyons or two photons. Because of the complexity of the expressions for the amplitudes, in particular the photon amplitude $\mathcal{A}^\gamma$, the computation time grows rapidly, limiting the laptop calculations to  $N\leq 12$.\footnote{In \cite{Rosenhaus:2021PRL,Gross:2021gsj,Bianchi:2023uby,Firrotta:2023wem} much larger $N$ values were considered, however the calculations in these papers were performed for the three-leg and four-leg processes including one HES state rather than two (IN and OUT) as in our case, and only for a subset of amplitudes, not for the whole S-matrix. In \cite{Hashimoto:2022bll} the $2\to 2$ process was considered as in our Eq.~(\ref{process}) but again only for individual amplitudes, considering $N\leq 27$. Essentially, calculating individual amplitudes instead of the S-matrix allows one to go to much higher levels. In \cite{Das:2023cdn}, where the whole S-matrix was calculated, the calculations was limited to modest values of $N$ just as in our case.} We first calculate the amplitudes for fixed kinematics and all possible partitions specifying the two HES states and from these we obtain the S-matrix on the subspace of fixed kinematics (which we call simply the S-matrix).

\subsection{The partition basis}\label{secchaospart}

The S-matrix is characterized by its eigenvalues and eigenvectors. Being unitary, it has eigenvalues on the unit circle, and of particular relevance for the RMT statistics are the eigenphases, i.e. the phases of the eigenvalues. The natural basis for the analysis of eigenvectors at HES level $N$ is the basis of partitions $\lbrace\mathcal{P}\rbrace$, i.e. the sets of $K$ numbers $\lbrace n_1,\ldots n_K\rbrace$ with $\sum_{k=1}^Kn_k=N$ and $K$ the highest nonzero occupation number. The dimension of the Hilbert space is the total number of partitions of $N$ denoted by $p(N)$.\footnote{As we know, this number grows exponentially as $p(N)\sim\exp\left(\sqrt{N}\right)$.}

We define the \emph{partition length} $\vert\mathcal{P}\vert$ as the \emph{total number of nonzero occupation numbers} $n_k$ in the partition (obviously between $1$ and $\mathrm{min}(N,K)$):\footnote{One might be confused that we denote partition elements, i.e. occupation numbers by $n_k$ whereas the DDF photon momenta were previously denoted by $N_kq$, where again $N_k$ are the elements of the partition. The reason is that $n_k=N_k$ only for \emph{open} strings. A closed string in the state $\vert n_k,\tilde{n}_k\rangle$ will have a different momentum, i.e. different $N_k$ even though the partition elements are the same. For this reason we keep the notation separate.}
\be 
\forall~\mathcal{P}=\lbrace n_1,\ldots n_K\rbrace\textrm{ with }\sum_{k=1}^Kn_k=N,~K=\mathrm{max~}j\vert_{n_j>0}:~~~\vert\mathcal{P}\vert\equiv\sum_j\Theta(n_j).\label{partlength}
\ee
In the above $\Theta(n_j)$ is the Heaviside unit step: $\Theta(n_j)=1$ for $n_j>0$ and $\Theta(n_j)=0$ otherwise. Now we can order all possible partitions according to partition lengths $\vert\mathcal{P}\vert$, from $1$ (the shortest partitions, of length $1$) to $N$ (the longest partitions, of length $N$); when several partitions have the same length their relative order is chosen arbitrarily. A more widespread way to characterize the size of the partition is the partition rank, defined as $m(\mathcal{P}_N)=\mathrm{max~}n_j\vert_{n_j>0}-\vert\mathcal{P}\vert$. While the rank lies between $-(N-1)$ and $N-1$, the length is between $1$ and $N$ and is thus more convenient for our purposes as we will also consider the logarithms of the lengths. 

A general eigenvector of the S-matrix is some linear combination of various partitions $\sum_{j=1}^{p(N)}c_j\vert\mathcal{P}_j\rangle$. We will argue that the distribution of the coefficients $c_j$ as a function of $j$ is a useful quantity when analyzing the scattering, in particular the relative contribution (i.e. the absolute value of the coefficients $\vert c_j\vert$) of short versus long partitions.

\subsection{Tests of the RMT statistics}\label{secchaosrmt}

Once we have the eigenvalues (and eigenvectors), we can apply the textbook tests of quantum chaos by comparing the eigenvalue statistics with the predictions for Gaussian random ensembles. As we mentioned in the Introduction, the RMT theory of chaotic scattering essentially replaces the eigenvalue spacings of the Hamiltonian in a closed system by the eigenphase spacings of the scattering matrix \cite{Smilansky1992}. To remind, eigenphase spacing is the difference between the phases of two neighboring eigenvalues:
\be
s_n\equiv\frac{\mathrm{arg~}\lambda_{n+1}-\mathrm{arg~}\lambda_n}{\bar{s}},\label{defrmta}
\ee
where $\bar{s}$ is the average value of the spacings for the whole S-matrix, and $\lambda_n$ are the eigenvalues ordered according to their phases, i.e. $\mathrm{arg}\lambda_1>\mathrm{arg}\lambda_2>\ldots>\mathrm{arg}\lambda_{p(\mathcal{N})}$ (or $\lambda_{p(\mathcal{N})^2}$ for the closed string). Therefore, \emph{if the scattering is chaotic in the RMT sense}, then the distribution of normalized spacings $s$ obeys the Wigner-Dyson distribution:
\be
P_\beta(s)=A_\beta s^\beta e^{-B_\beta s^2},\label{wd}
\ee
where $\beta=1,2,4$ for orthogonal (GOE), unitary (GUE) or symplectic (GSE) ensemble, respectively, and the constants $A_\beta$ and $B_\beta$ read
\bea 
A_1&=&\frac{\pi}{2},~B_1=\frac{\pi}{4}\label{wd1}\\
A_2&=&\frac{8}{\pi},~B_2=\frac{4}{\pi}\label{wd2}\\
A_4&=&\left(\frac{64}{9\pi}\right)^3,~B_4=\frac{64}{9\pi}.\label{wd4}\\
\eea
As we know, the three classes are determined by the properties under the time-reversal symmetry \cite{Haake1991} -- GOE describes time-reversal-invariant systems with "conventional time reversal operator", squaring to $1$; GSE describes time-reversal-invariant systems with Krammers degeneracy where time reversal squares to $-1$; GUE describes systems which break time-reversal invariance. Our process is time-reversal invariant and the strings are bosonic hence there cannot be any Krammers degeneracy. We thus expect the GOE statistics, with $\beta=1$. For purposes of checking the Wigner-Dyson statistics however we have tried fits with different $\beta$ values and we have also tried fitting the eigenphase spacing distribution with $\beta$ as a free fitting parameter. Finally, a regular scattering process can be written as a collection of independent channels, leading to the Poisson distribution of eigenphase spacings: $P_0(s)=\exp(-s)$. The crucial physical difference between the two is that the Wigner-Dyson distribution encapsulates eigenstate repulsion - one always has $P_\beta(s=0)=0$ and there is no clustering. The Poisson distribution, on the contrary, always has $P_0(s=0)=1$ and always leads to clustering of levels.

We will find that our S-matrices are typically not Gaussian random and certainly not Poissonian, but can -- in the simplest picture -- be thought of as being generated from a mixed ensemble with both Gaussian random matrices and integrable matrices. Therefore, it will make sense to consider the distribution
\be 
P_\mathrm{c}(s)\equiv w_PP_0(s)+(1-w_P)P_\beta(s).\label{wdpoisson}
\ee
Importantly, while both the Wigner-Dyson and the Poison form follow from rigorous derivations, the above linear combination is purely phenomenological -- there is no guarantee that a system with mixed (regular-chaotic) dynamics will indeed have Eq.~(\ref{wdpoisson}) for the eigenphase spacing distribution. We will use the size of $w_P$, i.e. the relative contribution of the Poisson contribution, as a rough indicator of non-chaoticity but it is not simply related to any physical quantity.

One important caveat concerning the Wigner-Dyson test is that the eigenstates with additional discrete symmetries that do not commute with time reversal can introduce novel behavior, leading to the Altland-Zirnbauer tenfold way instead of the Dyson threefold way \cite{Altland1996NonstandardSC}. In our case however there are no obvious additional symmetries that behave nontrivially under time reversal, so the threefold way of Eqs.~(\ref{wd1}-\ref{wd4}) remains valid.

Another measure of chaos which is perhaps more robust is the eigenspacing \emph{ratio}, which considers the ratio of the spacings between two neighboring pairs of eigenphases, i.e. between two neighboring spacings:
\be 
r_n\equiv\frac{s_n}{s_{n-1}}=\frac{\mathrm{arg~}\lambda_{n+1}-\mathrm{arg~}\lambda_n}{\mathrm{arg~}\lambda_{n}-\mathrm{arg~}\lambda_{n-1}}.\label{defrmtr}
\ee
Another possibility is to consider normalized ratios, defined as $R_n=\mathrm{min}(r_n,1/r_n)$. This chaos diagnostic was proposed in \cite{Oganesyan:2007PhysRevB.75.155111} and applied to field-theoretical systems in \cite{Srdinsek:2020bpq} and finally to HES scattering amplitudes in \cite{Bianchi:2023uby,Firrotta:2023wem}. Its appeal lies in the fact that the normalization cancels out hence this quantity is more robust to numerical fluctuations; in addition, it is argued in \cite{Bianchi:2023uby} to be less sensitive to the mixing of states with different symmetries. One can study the distributions $P(r)$, $P(R)$ or just the mean values $\langle r\rangle$, $\langle R\rangle$; the latter we find more useful as it gives robust and concise information within just a single number.

A few words about the calculations are in order. For smaller excitation numbers $N\leq 12$ it is easy to compute the S-matrix on a laptop in Wolfram Mathematica. For larger $N$ we perform the calculations on a cluster, and in addition we use approximate interpolation formulas for beta and gamma functions to speed up the calculation, sacrificing some accuracy. Still, when computing the whole S-matrix, we cannot go further than $N\approx 30$. This is an inherent difficulty when working with the S-matrix instead of just the amplitudes: the matrix becomes huge.

Now we are ready to study the properties considered in this section -- the structure of the S-matrix and its eigenvectors, and the eigenvalue statistics -- for the S-matrices of the HES-tachyon, HES-photon and HES-graviton scattering.

\section{S-matrix analysis of HES scattering processes}\label{secnum}

In this section we present numerical results on the S-matrix involving the HES states for the setting described in the previous section. We start with the open string and then move to discuss the physically more relevant closed string. Afterwards we provide some \emph{a posteriori} analytic arguments explaining the numerical findings.

\subsection{The HES-tachyon S-matrix}\label{secnumtach}

\subsubsection{The eigenphase statistics}\label{secnumtachrmt}

\begin{figure}[htb!]
\centering
(A)\includegraphics[width=0.95\linewidth]{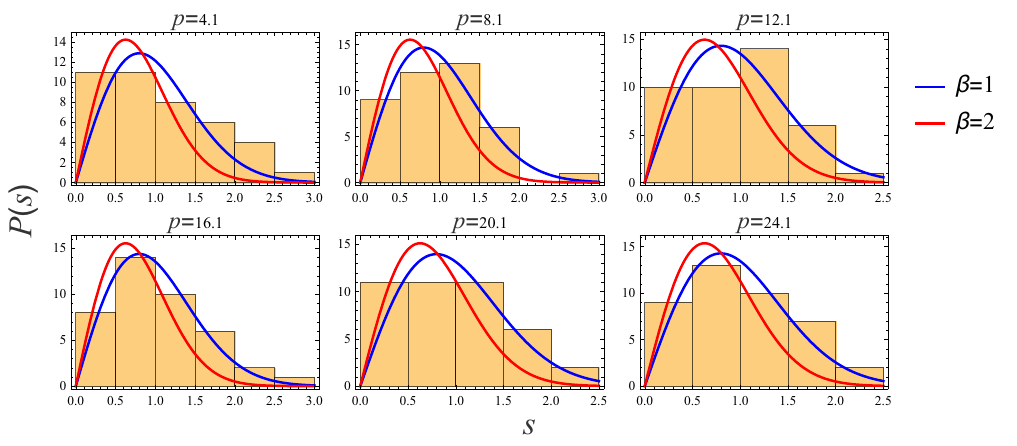}
(B)\includegraphics[width=0.95\linewidth]{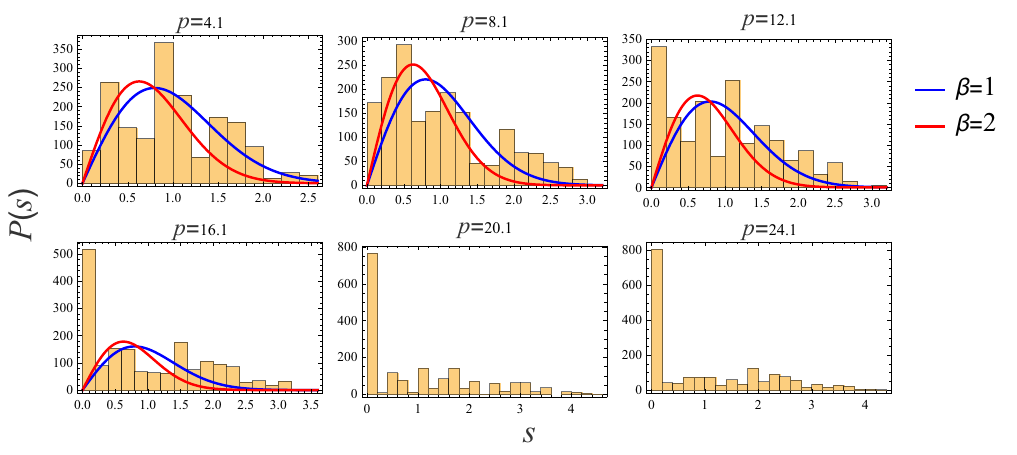}
%(A)\includegraphics[width=0.95\linewidth]{RMTOpen.pdf}
%(B)\includegraphics[width=0.95\linewidth]{RMTClosed.pdf}
\caption{Normalized eigenphase spacing distribution for six momentum values $p=4.1,8.1,12.1,16.1,20.1,24.1$, with $N = 10$, $\theta = 0.23$, $\theta' = 0.5$ and $\phi' = 0.7$, for open (A) and closed (B) HES. The blue and magenta lines represent the best fits to the Wigner-Dyson distribution (Eq.~\ref{wd}) with $\beta=1$ (Eq.~\ref{wd1}) and $\beta=2$ (Eq.~\ref{wd2}) respectively; the former corresponds to GOE, expected from the time-reversal symmetry and the latter to GUE, for time-reversal-breaking systems. While the overall shape of the distribution is close to the Wigner-Dyson function for intermediate momenta (roughly $p=16.1$ in (A) and $p=8.1$ in (B)), the presence of near-zero spacings clearly excludes pure RMT statistics. For the closed string there is strong clustering of levels at large momentum values; for that reason we do not give the fit to Wigner-Dyson distributions for the closed string for $p=20.1$ and $p=24.1$.}
\label{figrmtopenclosed}
\end{figure}

The first test that gives an idea on the degree of chaos in the S-matrix is the fit of the Wigner-Dyson distribution to the histogram of eigenphase spacings. Computing the S-matrix, diagonalizing it and producing the histogram of the differences of the eigenphases normalized by the mean of this difference, we typically obtain a picture such as Fig.~\ref{figrmtopenclosed}. We give the fit for both open and closed strings (panels (A) and (B) respectively), with the expected GOE ensemble value $\beta=1$, and with the GUE ensemble with $\beta=2$; we do not give the fit with $\beta=4$ as this case gives far worse fit quality (and is physically untenable in a bosonic system). The agreement with the RMT statistics is not impressive: in some cases (mainly $p=16.1$ for open strings and $p=4.1,8.1$ for closed strings) the agreement is decent but far from perfect. For other momenta, in particular for large momenta in the closed HES case, the disagreement is drastic. Indeed, at large momenta many eigenvalues coalesce and the spacings cluster at $s=0$. 

The natural conclusion is that dynamics is mixed, consisting of contributions from both Poisson and Wigner-Dyson distribution. Of course, for $N=10$, the value used in Fig.~\ref{figrmtopenclosed}, the set of spacings available for an open string is rather small, so the outcome may well be influenced by finite-size fluctuations. But the closed string already has a much larger S-matrix (of size $p(N)^2\times p(N)^2$ instead of $p(N)\times p(N)$) and significant finite-size effects seem unlikely.

%Let us first apply the basic tests for the eigenphase spacing and spacing ratio distributions and mean values. Typical results for both open and closed strings are given in Fig.~\ref{figrmtopenclosed}, together with the fit to the GOE and GUE prediction (Eqs.~\ref{wd1}-\ref{wd2}). The overall shape of the histograms is relatively close to the Wigner-Dyson distribution (\ref{wd}), in particular for $\beta=1$ as one should expect from time-reversal symmetry. RMT implies strict eigenphase repulsion and the presence of near-zero spacings, \emph{even if the overall shape of the distribution is close to the Wigner-Dyson function}, means that the underlying physics \emph{cannot} be of RMT type. We can thus tentatively conclude that the HES scattering exhibits a mixed state space, with both regular (Poisson) and chaotic (Wigner-Dyson) channels. 

\begin{figure}
\centering
(A)\includegraphics[width=0.45\linewidth]{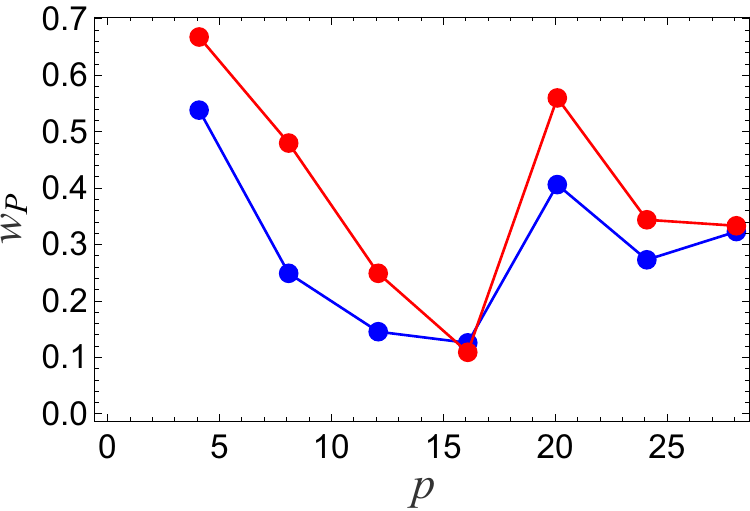}
(B)\includegraphics[width=0.45\linewidth]{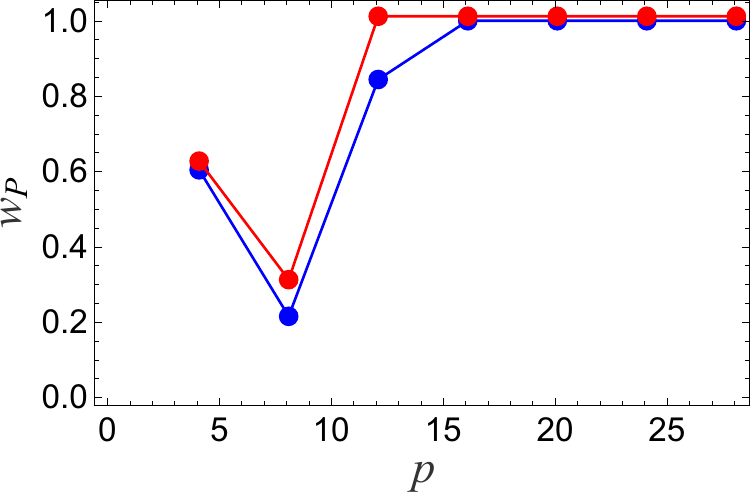}
\caption{The relative contribution of the Poisson distribution $w_P$ when fitting the Poisson-Wigner-Dyson combination to the histogram of eigenphase spacings as a function of momentum, for the same values of momenta and angles as in Fig.\ref{figrmtopenclosed}, for open (A) and closed (B) string. The smallest regular component, i.e. the smallest Poisson ratio $w_P$ ($0\leq w_P\leq 1$) is found for $p\approx 16.1$ (A) and $p\approx 8.1$ (B), roughly the same values which yield the nicest fit to the Wigner-Dyson distribution in Fig.~\ref{figrmtopenclosed}. For large momenta we again see that the closed string becomes completely regular, with strong clustering of eigenphases. We fit the Poisson+Wigner-Dyson distribution both with $\beta=1$ (blue) and $\beta=2$ (red). The GOE value $\beta=1$ gives somewhat better plots as we expect. The full lines are just to guide the eye.}
\label{figwpopenclosed}
\end{figure}

The hypothesis of two-component dynamics (regular and chaotic) is checked by fitting the combination of the two distribution functions (Eq.~\ref{wdpoisson}). The relevant quantity now is $w_P$, the contribution of the Poisson distribution, which is expected to be minimal when chaos is the strongest. Its dependence on the momentum $p$ is given in Fig.~\ref{figwpopenclosed}. Here and in most other figures throughout the paper we perform the fit both with $\beta=1$ and $\beta=2$ but in general the former value, corresponding to GOE and time-reversal invariance agrees better with the data as it should be.

\begin{figure}
\centering
(A)\includegraphics[width=0.45\linewidth]{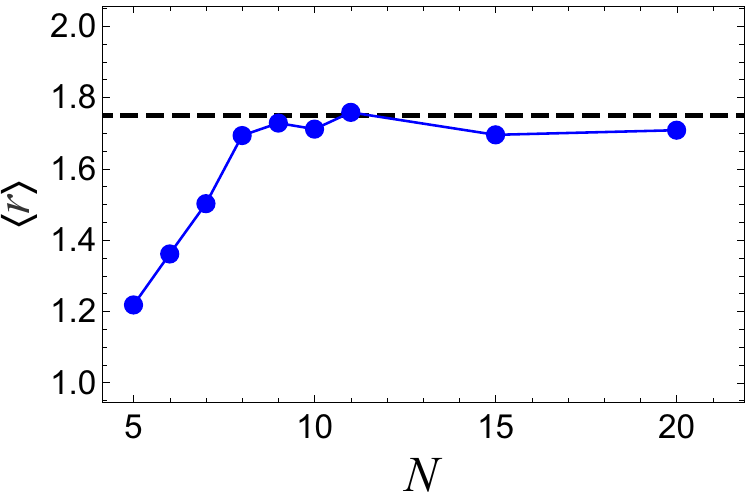}
(B)\includegraphics[width=0.45\linewidth]{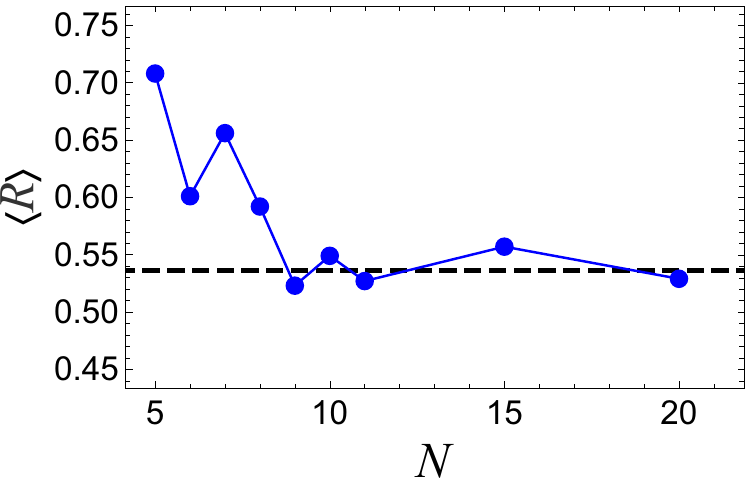}
\caption{The non-normalized (A) and normalized (B) average spacing ratio at the crossover momentum $p_c=16.1$ for the closed HES-tachyon scattering at different excitation levels $N$ of the HES; the kinematic parameters are the same as in Figs.~\ref{figrmtopenclosed} and \ref{figNropenclosed}. For growing $N$ the spacing ratios stay firmly around the values $\langle r\rangle=1.75$ (A) and $\langle R\rangle=0.536$ (B) predicted for the GOE random matrix dynamics (black dashed reference lines). The spacing ratio indicates stronger chaos at $p\approx p_c$ than the fits to the Wigner-Dyson distribution, which even around $p_c$ show visible discrepancies.}
\label{figNropenclosed}
\end{figure}

Another way to characterize the proximity to the RMT-like chaotic behavior is the average spacing ratio $\langle r\rangle$ defined in Eq.~(\ref{defrmtr}), or its normalized variant $\langle R\rangle$. We find it particularly convenient when studying the behavior of the S-matrix for increasing excitation number $N$. We have tentatively found that the scattering shows clear signs of chaos only for certain values of momentum, which we dub the crossover momentum $p_c$ for reasons that will soon become clear. Now we fix this value ($p_c=8.1$ for the open string and $p_c=16.1$ for the closed string) and vary the total excitation number $N$ (Fig.~\ref{figNropenclosed}). Computing $\langle r\rangle$ (A) and $\langle R\rangle$ (B) for each $N$ value at $p=p_c$ we find that already for $N\gtrsim 9$ the values are very close to the predicted outcome for the time-reversal-invariant GOE random matrix ensemble (black dashed line in Fig.~\ref{figNropenclosed}). This suggests that indeed in a small interval of momenta around $p_c$ the S-matrix shows clear signs of chaos which stay stable for large $N$ (in accordance with the string/black hole complementarity and the fast scrambling of black holes, if there is strong chaos then indeed it has to remain strong (or become stronger and stronger) as $N$ grows). Still, a more refined measure such as the fit to the Wigner-Dyson distribution or its combination with the Poisson distribution (Figs.~\ref{figrmtopenclosed},\ref{figwpopenclosed}) reveals that even at $p_c$ some non-chaoticity remains. Therefore, rather than a sharp strongly chaotic point it is a smeared crossover region where chaos becomes strong but still non-uniform. Now we will try to understand the nature of this crossover.

\subsubsection{The crossover}\label{secnumtachcross}

We now invoke the partition basis described in the subsection \ref{secchaospart} and the notion of partition length from Eq.~(\ref{partlength}). Finding the coordinates $c_k^{(n)}$ ($1\leq k\leq p(N)$ for open HES and $1\leq k\leq p(N)^2$ for closed HES) of the $n$-th S-matrix eigenvector $\vert n\rangle$ in the partition basis, we can speak of the relative contribution of short vs. long partitions to the eigenvector (large $c_k^{(n)}$ for small/large $k$ implies dominant contribution of short/long partitions). One detail is still arbitrary: the ordering of the eigenvectors themselves, i.e. the number $n=1,\ldots p(N)$ ($n=1,\ldots p(N)^2$ for closed strings) in $c_k^{(n)}$ as defined above. We opt to order the eigenvectors according to the absolute value of the real part of their eigenvalues: thus $\vert 1\rangle$ is the "leading" eigenvector, which contributes most to the OUT state after the scattering, and $\vert p(N)\rangle$ ($\vert p(N)^2\rangle$ for closed strings) is the "least important" eigenvector, which contributes the least.

\begin{figure}
\centering
(A)\includegraphics[width=0.95\linewidth]{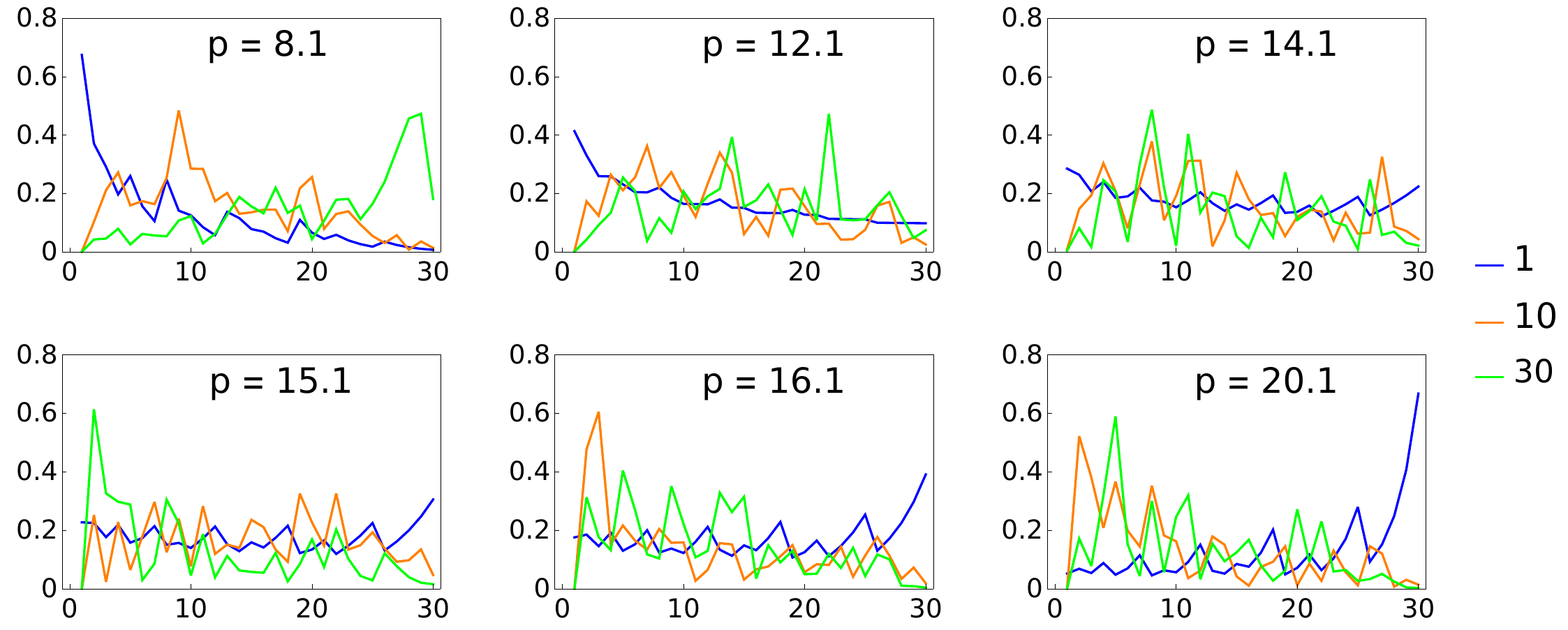}
(B)\includegraphics[width=1.05\linewidth]{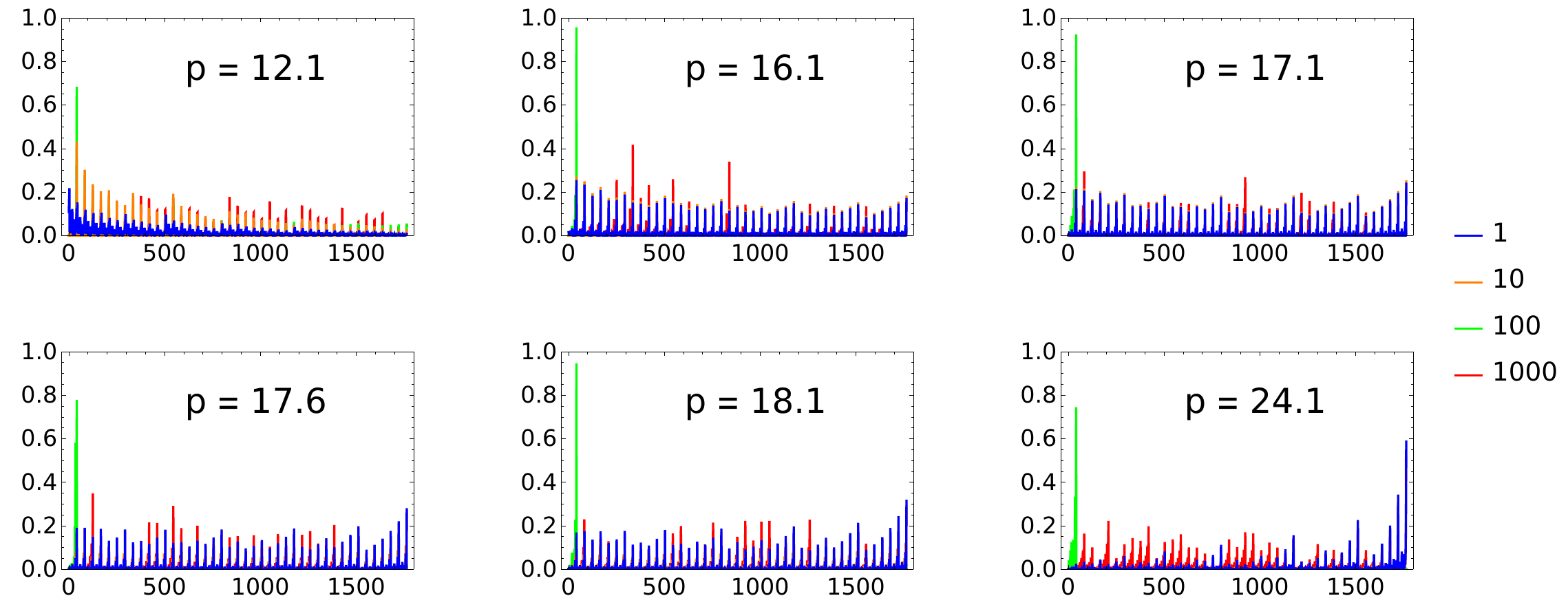}
\caption{Coefficients of the S-matrix eigenvector $\vert n\rangle$ in the partition basis, for the leading eigenvector and a few smaller eigenvectors, denoted by colored lines (see the plot legends). The partitions are enumerated from shortest (of the form $(0,\ldots 0,N,0,\ldots 0)$) to longest (of the form $(1,\ldots 1)$). The leading eigenstate $n=1$ (blue line -- corresponding to the largest eigenvalue, which dominates the final state of the scattering process) consists mainly of short partitions for small momenta, becomes approximately equipartitioned for intermediate momenta which we identify with the crossover scale $p_c$, and consists mainly of long partitions for high momenta. The eigenstates with $n=10$ and $n=30$ behave in a more or less complementary way (i.e. long partitions dominate for $p<p_c$ and short partitions for $p>p_c$), although for them the trends are less clear. Six momentum values are considered ($p=8.1,12.1,14.1,15.1,16.1,20.1$ in (A), for open strings, and $p=12.1,16.1,17.1,17.6,18.1,24.1$ in (B), for closed strings), with $N = 9$, $\theta = 0.23$, $\theta' = 0.30$, $\phi' = 0.70$ (A) vs. $N = 10$, $\theta = 0.20$, $\theta' = 0.30$ and $\phi' = 0.20$ (B). The crossovers in general happen at different momenta for open and closed strings; here we have $p_c\approx 15.1$ for open HES (A) and $p_c\approx 17.5$ for closed HES (B).}
\label{figpartopenclosed}
\end{figure}

In Fig.~\ref{figpartopenclosed}, we plot the set of coordinates $c_k^{(n)}$ for three eigenvectors, for a number of S-matrices with different momenta and fixed angles $\theta, \theta'$ and $\phi'$. We find that short partitions dominantly contribute to the leading eigenvector (the one with the largest eigenvalue) at small momenta $p$, while for large momenta long partitions dominate. The crossover from the domination of short partitions to the domination of long partitions happens just around the momentum where chaos is maximal, which we have denoted by $p_c$ in the previous subsection: at $p\approx p_c$ partitions of all lengths contribute equally. We find that $p_c$ slightly decreases with increasing $\phi'$ but overall is almost insensitive to the kinematics (for the tachyonic case that we consider here).

\begin{figure}
\centering
(A)\includegraphics[width=0.9\linewidth]{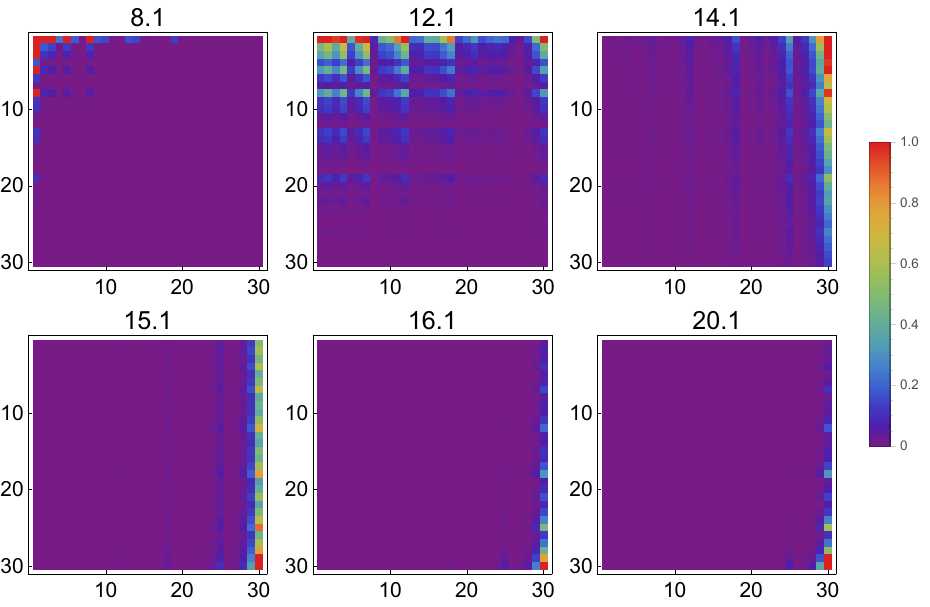}
(B)\includegraphics[width=0.9\linewidth]{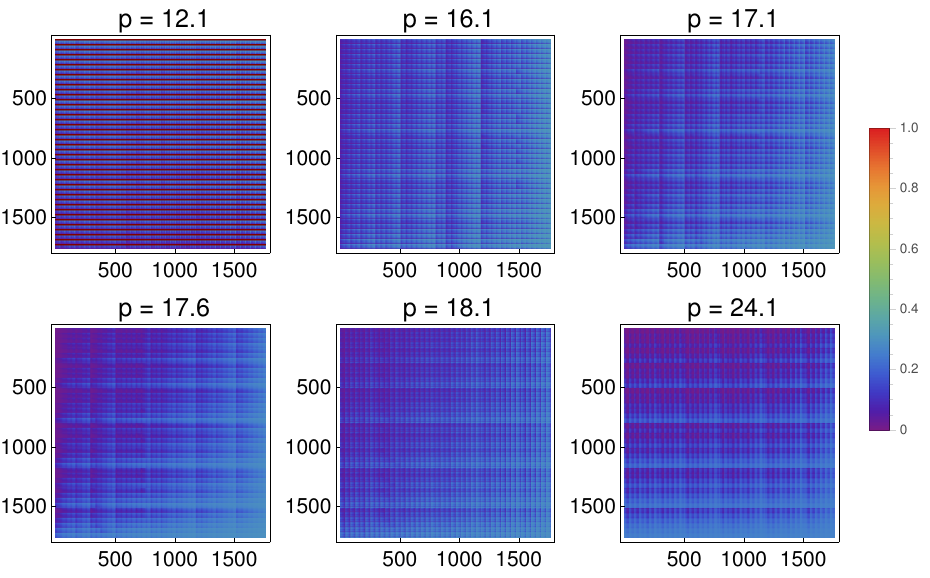}
\caption{The S-matrices in the partition basis for open (A) and closed (B) HES scattering, for the same kinematics as in Fig.~\ref{figpartopenclosed}, the color code showing the module of the (generically complex) matrix elements. For the open string the S-matrix and consequently the interval of the values of its elements is much smaller hence it is convenient to use a linear color scale (normalized to the unit interval). For the closed string the S-matrix is much larger and the magnitude of its elements varies by several orders of magnitude hence we use the logarithmic color scale (again normalized to the unit interval). In accordance with Fig.~\ref{figpartopenclosed}, the dominant processes for $p<p_c\approx 15$ involve short partitions with few occupied levels and large occupation numbers (upper left corner), whereas for $p>p_c$ the opposite is true and the lower left corner of the S-matrix is dominant.}
\label{figsmatopenclosed}
\end{figure}

The fact that at large energies (and momenta) long partitions dominantly contribute to the dynamics is also seen from the plot of absolute values of the S-matrix elements in Fig.~\ref{figsmatopenclosed} (remember the rows of the matrix are nothing but the eigenvectors in the partition basis): the largest matrix element migrates from upper left corner (processes scattering short states into short states) to upper right corner (short-to-long processes) to bottom right corner (long-to-long processes).\footnote{This was also seen and discussed in the recent paper \cite{Das:2023cdn}.} This can be roughly understood in a simple way: for $p\ll p_c$ there is not enough energy to activate most modes so only a few modes contribute (and they must have large occupation numbers $n_k$ in order to have the total excitation number $N$); for $p\gg p_c$ the kinetic energy is much larger than the interaction energy scale thus all the modes are excited (and they must have mostly small occupation numbers so as not to overshoot the total occupation $N$). For $p\sim p_c$ these two factors balance each other and the dominant eigenvectors consist of partitions of all lengths.

Now why is the chaos clearly visible precisely around these crossover momenta $p_c$? For small energies only a few states effectively participate in the scattering dynamics hence it cannot be very complex. As we increase the energy more and more modes are activated providing more channels for the interaction, forming a complex structure through which chaotic behavior may develop. But at very high energies there is "less time" for the interaction to occur, the strings just "fly away from each other", which results in the suppression of the chaotic behavior.\footnote{This situation is analogous to classical scattering, where likewise at low energies the skeleton of periodic orbits (which defines the symbolic dynamics) is quite simple as most orbits see just a near-harmonic potential well, at high energies most orbits barely feel the scattering potential and continue to infinity almost undisturbed, and at intermediate energies the competition between the bounded and unbounded dynamics generates sensitivity to initial conditions and chaos.}

\begin{figure}
\centering
\includegraphics[width=0.95\linewidth]{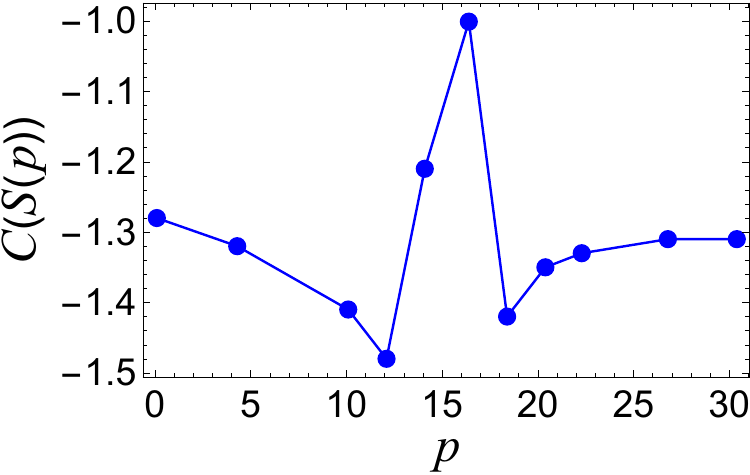}
\caption{Shannon entropy $C(S(p))$ in computational units as a function of momentum $p$ for the same kinematics as in Fig.~\ref{figpartopenclosed}(B). We observe a peak in the entropy at the crossover momentum $p_c \approx 17.5$. The line is just to guide the eye.}
\label{figent}
\end{figure}

We further support the above reasoning by calculating the Shannon information entropy associated with the S-matrix. As we know, the Shannon entropy is defined as $\mathcal{S}=\sum_jp_j\log p_j$ for some classical probabilistic system with probability $p_j$ assigned to each state $j$. The usual quantum-mechanical analogue is the von Neumann entropy, defined in terms of the density matrix $\rho$ as $\mathrm{Tr}\rho\log\rho$. Since we have a quanutm scattering process, one should in principle speak of von Neumann rather than Shannon entropy. However, we work with the S-matrix and do not have the density matrix so defining the von Neumann entropy the canonical way would be tricky. We can introduce a phenomenological measure of entropy or complexity for a single amplitude from the IN state $\vert\vec{n}\rangle$ to the OUT state $\vert\vec{m}\rangle$ simply as:
\be
C\left(\vert\vec{n}\rangle\to\vert\vec{m}\rangle\right)=\mathrm{Re}\,\left[\mathcal{A}_{\vert\vec{n}\rangle\to\vert\vec{m}\rangle}\log\mathcal{A}_{\vert\vec{n}\rangle\to\vert\vec{m}\rangle}\right].\label{entamp}
\ee
Taking the real part is necessary as amplitudes are in general neither real nor positive. For the whole scattering matrix, the natural choice is to sum the definition (\ref{entamp}) for all amplitudes, or equivalently to take the trace of the expression over the whole S-matrix. But in the eigenbasis the S-matrix becomes diagonal, with elements $\mathcal{N}\exp(i\phi_k)$, where $\phi_k$ is the eigenphase of the state $k$ and $\mathcal{N}$ is some overall normalization constant, which is irrelevant for our purposes. Therefore we can write
\be 
C(S)=\mathrm{Re}\,\mathrm{Tr}\,S\log S=-\sum_k\phi_k\sin\phi_k\label{entsmat},
\ee
where the sum goes from $k=1$ to $k=p(N)$ for open strings and from $k=1$ to $k=p(N)^2$ for closed strings.\footnote{For $N\to\infty$ the index of the state becomes continuous and we get the differential entropy, the well-known generalization of the Shannon entropy for continuous distributions. This limit is important for the string/BH complementarity but we leave it for further work.} As the S-matrix becomes chaotic, we expect the information entropy to increase, reflecting the increase in the complexity of dynamics. In fact, it is likely possible to calculate $C(S)$ in a closed form for a Gaussian random matrix ensemble but so far we have not found an expression in terms of elementary functions. Still, Fig.~\ref{figent} suggests that $C(S)$ is indeed a useful measure as it becomes maximal at about the same value where the spacing analysis and the eigenvector structure from Fig.~\ref{figpartopenclosed}(B) show the crossover and strong chaos.

\subsubsection{Quasi-invariant states}

\begin{figure}[htb!]
\centering
(A)\includegraphics[width=1.0\linewidth]{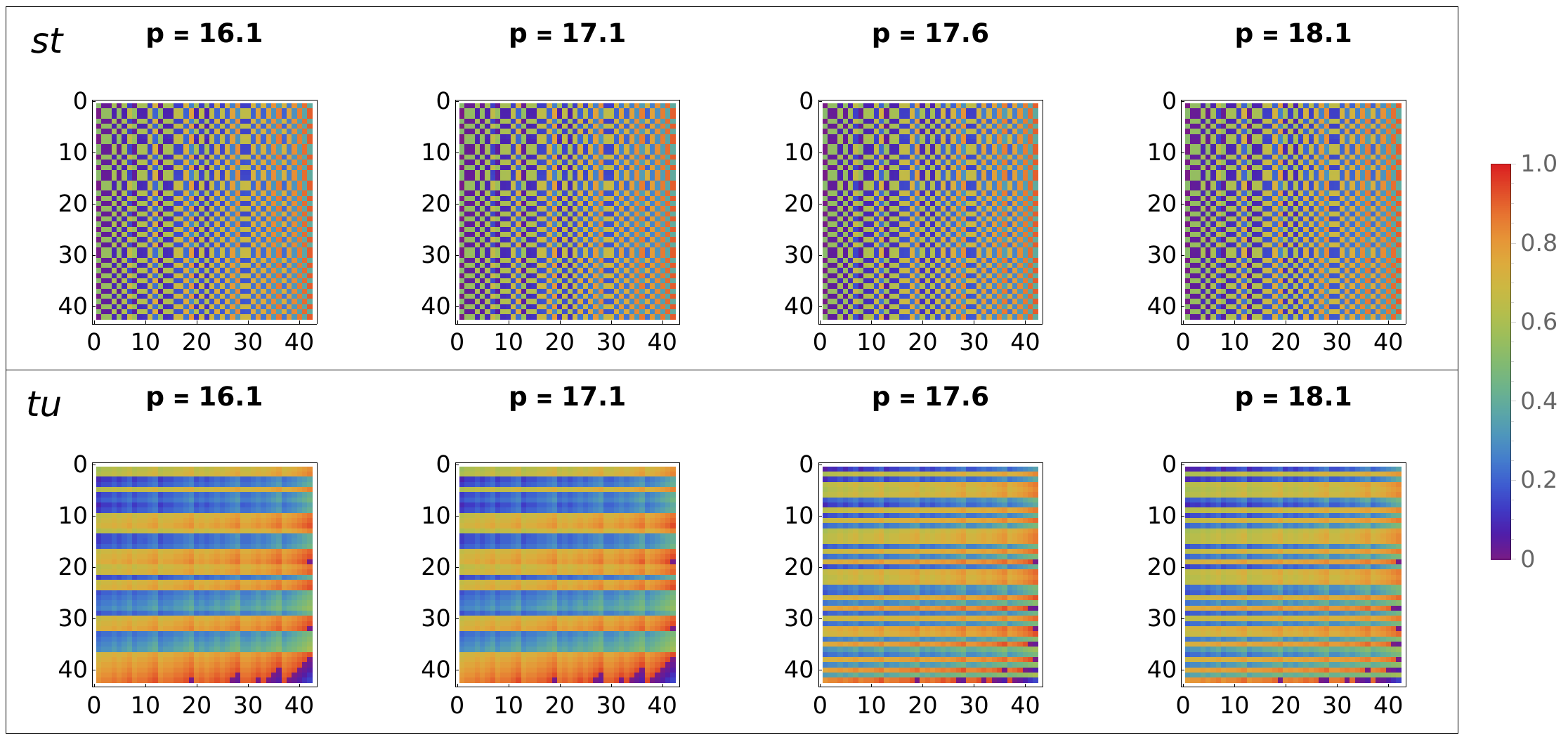}
(B)\includegraphics[width=1.0\linewidth]{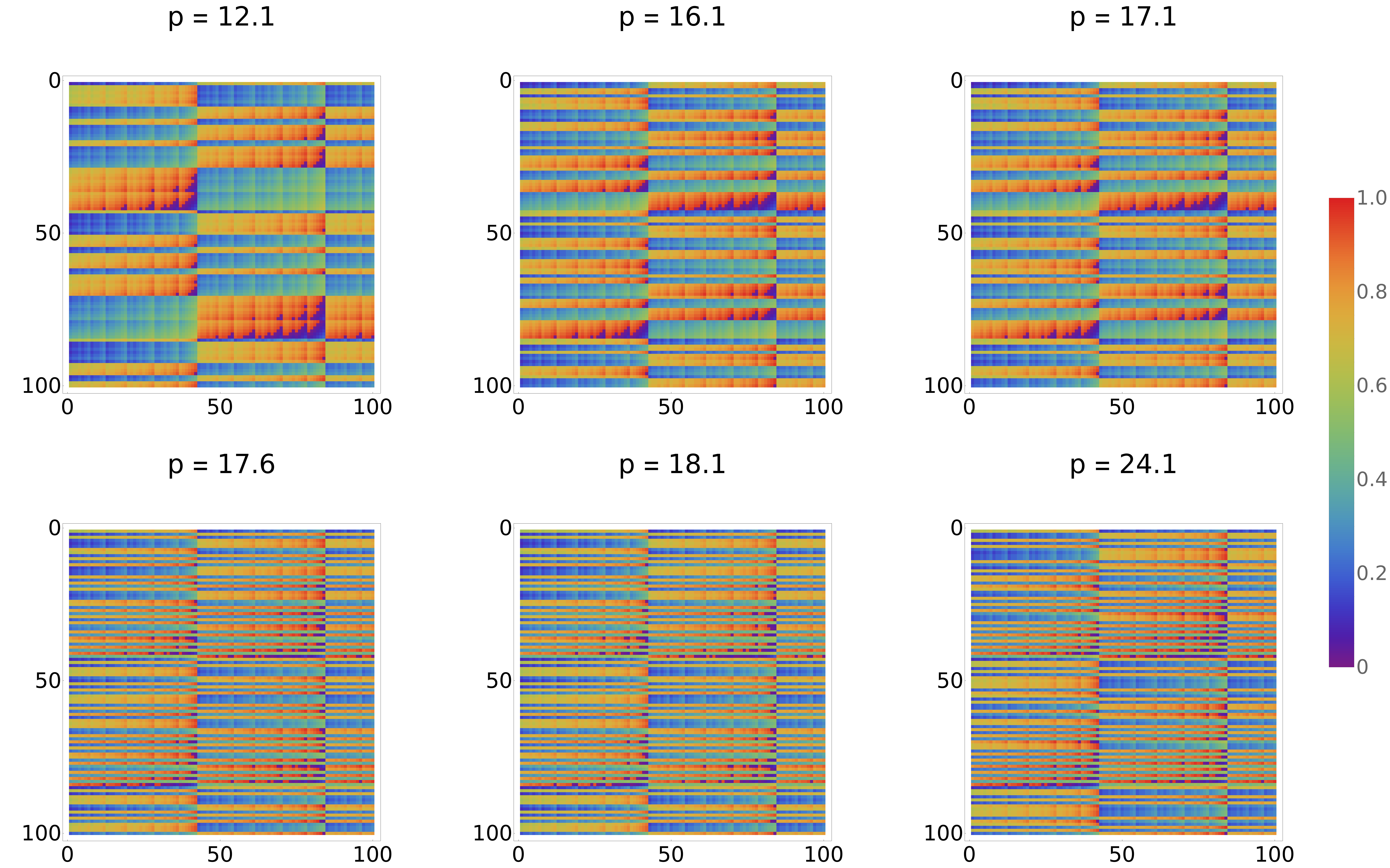}
\caption{Channels $\mathcal{A}_{st}$ and $\mathcal{A}_{tu}$ of the open string (A) and the closed string (B) amplitude in the partition basis for the same kinematics as in Fig.~\ref{figpartopenclosed}, the color code showing the phases of the elements, normalized to the $[0,1]$ interval. The $tu$ part shows stripes (i.e., sets of partitions with lengths confined to some fixed interval), whose number increases at the crossover. The phase structure of the $st$ part is on the other hand more or less featureless. In (B) only a part of the S-matrix is shown in order to zoom-in into the stripe structure; the rest of the S-matrix behaves in a similar manner.}
\label{figsttuopenclosed}
\end{figure}

Returning again to the structure of the S-matrix akin to plots in Fig.~\ref{figsmatopenclosed}, it is possible to detect the regular structure that spoils the Wigner-Dyson statistics at all momenta (to some extent even at $p_c$). For this it is necessary to plot different channels separately as they can behave very differently. In Fig.~\ref{figsttuopenclosed} we plot the $st$ and $tu$ parts of the open string amplitude. In the $tu$ matrix we see clear structures -- stripes, while in the $st$ part we do not see any obvious regularities. Of course, the stripe structure of the $tu$ channel induces similar structure in the complete S-matrix (though less sharp as it combines with the featureless structures of the other channels). 

These quasi-invariant states are very intriguing for several reasons: (1) they provide a specific mechanism which precludes the canonical random matrix behavior of the HES S-matrix -- we need to understand them if we want to know why the HES approach can seemingly never peep into the black hole regime (2) they are a natural generalization of the scar states which have drawn a lot of attention in quantum chaos studies in recent years but here we find quasi-invariant states \emph{generically}, not for fine-tuned initial conditions as is usually the case for scars\footnote{In the context of our calculation of first-quantized string dynamics, we may see something like quantum-mechanical scars introduced in \cite{Heller:1984zz}, rather than the scars in quantum many-body systems such as those in \cite{Moudgalya:2021xlu,Liska:2022vrd,Milekhin:2023was}.} (3) analyzing the structure of the amplitudes starting from the expressions (\ref{ast}-\ref{ckdef}) for the tachyon or (\ref{astphoton}-\ref{coeffbarbard2p}) for the photon it might be possible to understand the quasi-invariant states analytically in future work. For now we can only give some general hints as to their appearance, which we do in the discussion in Section \ref{secconc}.

\subsection{The HES-photon and HES-graviton S-matrices}\label{secnumphot}

For the HES-photon scattering (Eq.~\ref{processgamma}) and its closed-string equivalent, the HES-graviton scattering (Eq.~\ref{processgrav}) we can perform a similar analysis as for the HES-tachyon scattering discussed in the previous subsection. The overall phenomenology is similar, with two differences: 
\begin{enumerate}
    \item Dependence of dynamics on the momentum $p$ and the angles $\theta$ and $\theta'$ is now much stronger and more complicated, so that in some cases there is more than one value of momentum where chaos becomes strong, but it only does so when the values of the angles are also inside the chaotic window.
    \item The strong chaos found for special $p$ and $\theta$ values is mainly there for finite $N$ values and does not in general persist for growing $N$. 
\end{enumerate}

\begin{figure}
\centering
(A)\includegraphics[width=0.45\linewidth]{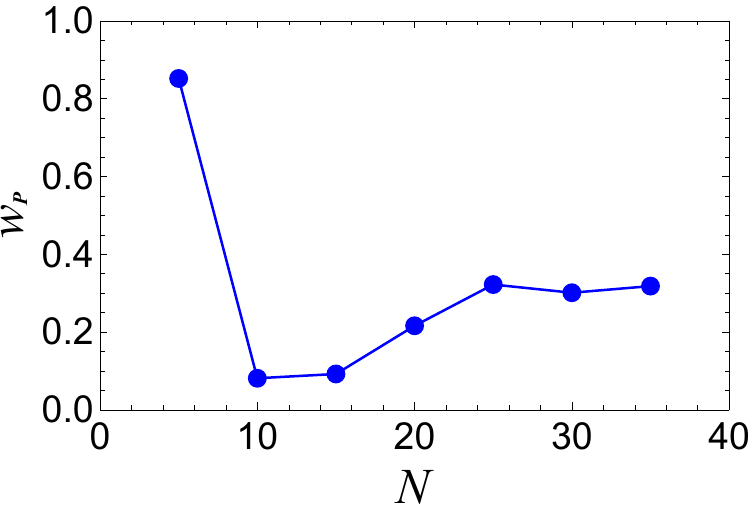}
(B)\includegraphics[width=0.45\linewidth]{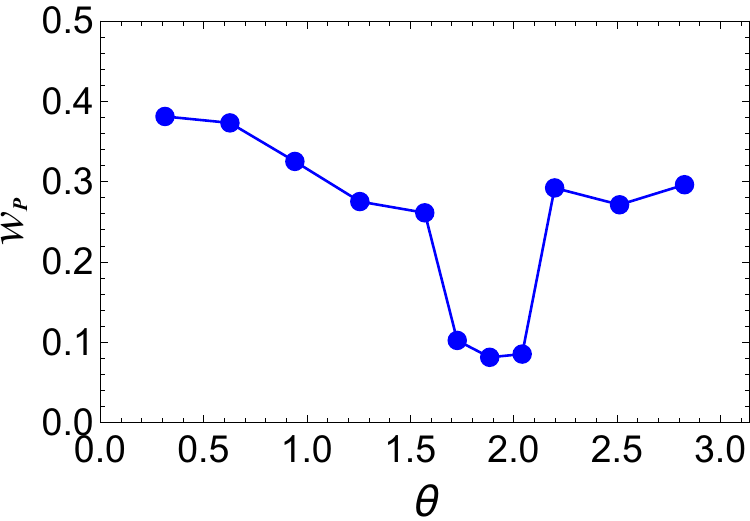}
\caption{The contribution of the Poisson distribution to the fit of the level spacing histogram of the HES-graviton S-matrix, as a function of $N$ for $\theta=0.70$ (A), and as a function of $\theta$ for $N=10$ (B), in both cases for $p=10.1$, $\theta'=0.25$, $\phi'=0.70$. There is obviously a relatively narrow interval of angles where the eigenphase spacing distribution is almost of the RMT type, but this is not the case for all values of $N$. For large $N$ values, when HES becomes really highly excited, the chaos is actually weaker. The fit is for $\beta=1$, and the solid lines are just to guide the eye.}
\label{figthetaphoton}
\end{figure}

The above findings can be seen from Fig.~\ref{figthetaphoton}. There is now a narrow interval in the angle $\theta$ where the Poisson contribution $w_P$ becomes very low but this is only the case for $N=10$ and $N=15$, otherwise the dynamics is again highly mixed. This suggests that the mechanisms of chaos at work here are not relevant for the black hole scrambling and the string/black hole complementarity picture: otherwise the chaos could only become stronger for growing $N$. Therefore, unlike the tachyon case considered in Fig.~\ref{figNropenclosed}, where indeed the strong chaos at $p\approx p_c$ persists for growing $N$, here it is a finite-$N$ phenomenon.

\begin{figure}
\centering
(A)\includegraphics[width=0.45\linewidth]{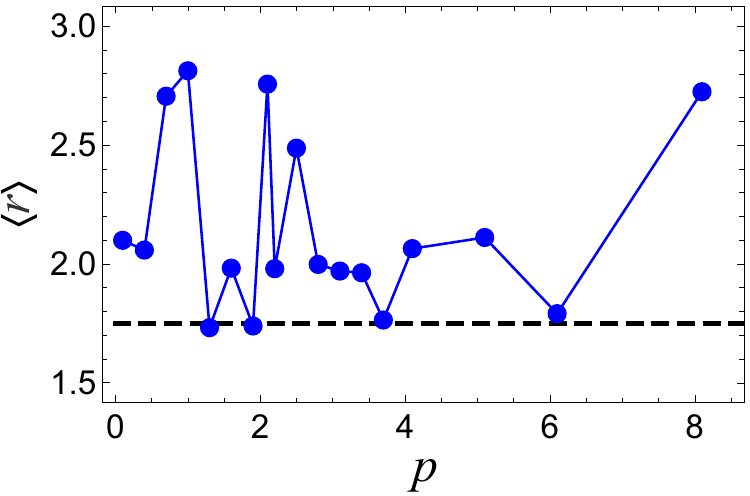}
(B)\includegraphics[width=0.45\linewidth]{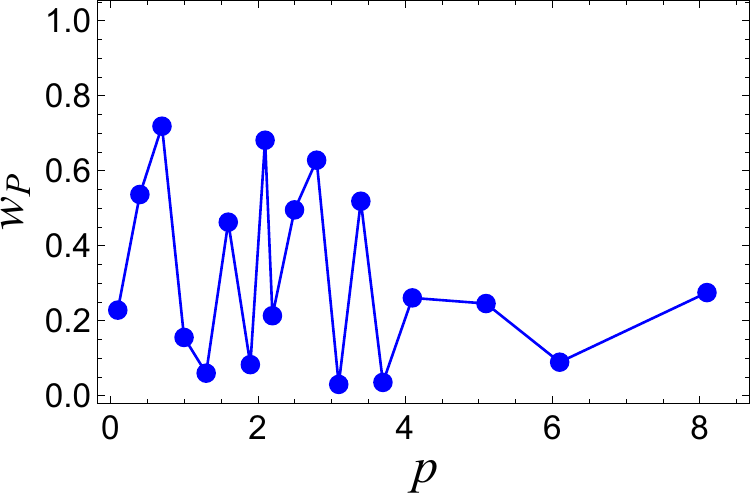}
\caption{Mean level spacing ratio $\langle r\rangle$ (A) and the Poisson contribution to the eigenphase spacing histogram $w_P$ for a HES-photon process with $N=9$, $\theta=0.23$, $\theta'=0.5$, $\phi'=0.7$, and a range of momenta $p$. Now we have several chaotic momenta instead of just one, roughly at $p_c=1.3,1.9,3.7,6.1$. Both measures of chaos ($\langle r\rangle$ and $w_P$) roughly agree in indicating chaotic dynamics at these points. The fit is for $\beta=1$ in the Wigner-Dyson distirbution.}
\label{figpphoton}
\end{figure}

\begin{figure}
\centering
\includegraphics[width=1.0\linewidth]{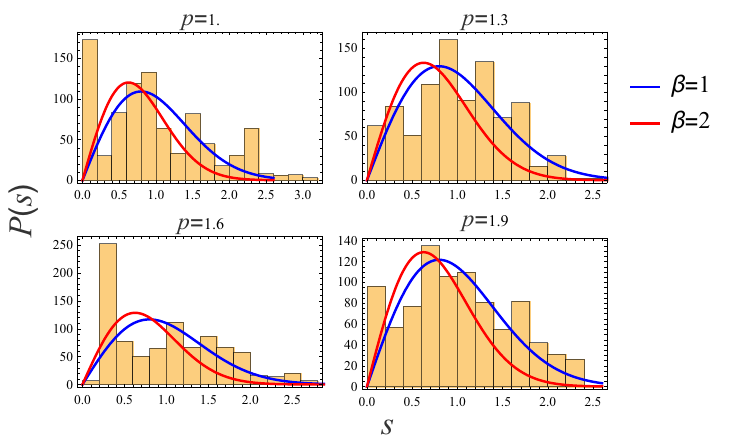}
\caption{Eigenphase spacing distribution for the same kinematics as in Fig.~\ref{figpphoton} for four momentum values $p=1.0,1.3,1.6,1.9$. The points $p=1.3$ and $p=1.9$ provide a better fit to the Wigner-Dyson distribution with $\beta=1$/$\beta=2$ (blue/red line) than the other two, in accordance with the other chaos indicators ($\langle r\rangle$ and $w_P$).}
\label{figrmtphoton}
\end{figure}

The momentum dependence of the chaos indicators $\langle r\rangle$ and $w_P$ can be seen in Fig.~\ref{figpphoton}. Now we have several crossover values $p_c$ where the main contribution to the eigenvectors goes back and forth between short and long partitions. In this example four $p_c$ values exhibit clear signs of chaos. We note in passing that for mixed/weakly chaotic cases the $\langle r\rangle$ values are now above the GOE value $1.75$ whereas for the tachyon they are below. We do not know the significance of this. For reference we also give a few plots of the eigenphase spacing distribution in Fig.~\ref{figrmtphoton}, for two crossover momenta ($p=1.3,1.9$) and two other momenta ($p=1.0,1.6$), where we can directly observe an increased fit quality by the Wigner-Dyson distribution with $\beta=1$ for the $p=p_c$ momenta. Here again we perform the Wigner-Dyson fit also for $\beta=2$ in addition to $\beta=1$ for comparison, but the fit is again better for $\beta=1$.

In conclusion, the scattering of spinful states (such as the photon) provides a more complex phenomenology: strong chaos is still limited to special values of parameters, not only momenta but also the angles and the occupation number. In this sense one could argue that this case is even less chaotic than the tachyonic case as chaos does not persist for infinite $N$. On the other hand, highly erratic dependence on $p$ and $\theta$ is more in accordance with the classical chaotic scattering where precisely the high sensitivity of the scattering outcome on the initial conditions encapsulates chaos \cite{Ott93,ott_2002}. In fact, \cite{Hashimoto:2016dfz} has found precisely such behavior also for the HES-tachyon process but for individual amplitudes only (here we have shown it does not hold for the S-matrix as a whole, where just a single $p$ value is special). The result we have is akin to the classical scattering of the ring string off a black hole, studied e.g. in \cite{Frolov1999,BasuAdSS,BasuRN} and numerous later works. 

\section{Discussion and conclusions}\label{secconc}

The S-matrices paint a complex picture of HES dynamics. The dynamics varies from mostly regular, with strong eigenphase clustering, to mostly chaotic, in good agreement with the usual RMT measures of chaos. There is no simple characterization of chaos as a function of kinematics or the occupation number $N$, and spinful vs. scalar incoming state also influences the outcome. This variation is in itself our first conclusion: the dynamics is strongly non-uniform and no signs of black hole universality are seen. 

The first important idea is that of a crossover at special momentum $p_c$ between two mostly regular regimes in the HES-tachyon scattering: only when all the HES states are excited in the scattering process the chaos dominates in the eigenphase statistics. For scattering of photons and gravitons the picture is more complicated, and several values of special/chaotic momenta and angles can be found. To some extent this is expected: only a non-scalar state can exhibit strong dependence on the angles and geometry of the scattering process. But more important is the very idea of the crossover. First, it tells us how the chaos arises: it arises through the exponentially large number of excited states, a factor which makes up for the simplicity of the Polyakov action for the scattering and the Veneziano- and Virasoro-type amplitudes.\footnote{Indeed, one may wonder how is it possible to see any chaos when the action is quadratic. The answer lies in the fact that the Polyakov action has an additional constraint and, as we mentioned, in the fact that we have an exponentially large number of amplitudes in the S-matrix, where each amplitude behaves as an infinite sum of field-theoretical amplitudes.}

The second important idea is that of quasi-invariant states. They provide a clear reason why chaos can never become uniform: as long as there are many IN states which give almost the same OUT state there will be a strong clustering of eigenphases and no uniform eigenphase repulsion. We plan to understand these states better in future work.

Both the crossover (mentioned also in \cite{Das:2023cdn}) and the quasi-invariant states could not be seen from amplitudes alone, they are properties of the whole S-matrix. Therefore it is important to move away from the study of individual amplitudes toward the study of the whole S-matrix. Yet, this is computationally demanding. Indeed, one possible weakness of this work is that we are forced to study relatively small $N$ values (roughly around $10$) when working on a laptop, and even on a cluster we cannot move beyond $N\approx 30$ and that with an approximate evaluation of the gamma and beta functions. Therefore, one might worry that we simply do not look at sufficiently high $N$ to make our strings truly highly excited. Could it be that in fact chaos will become uniform and black-hole-like when going for much higher excitation numbers?

We cannot provide a definite answer but we have some reasons to conjecture that the answer is negative. First, we reach a seemingly constant ratio of regular to chaotic (Poisson to Wigner-Dyson) eigenphase populations: it is possible but not very likely that this constant ratio will suddenly drop at some $N$. Also, when studying individual amplitudes the authors of \cite{Bianchi:2023uby} have gone for much higher $N$ (over $10^5$ in some cases) and they still do not see the statistics collapse onto the RMT predictions. Finally, existing results on the eigenvalue distribution of beta-Wishart matrices \cite{2013JMP....54h3507D,e18090342} provide a path toward an analytic proof that quasi-invariant states are present for S-matrices of arbitrarily large size. We plan to follow this path in further work on the subject.

All of the above shows that HES scattering is very different from black hole scrambling: in the latter, there are universal time and space scales encapsulated in the chaos bounds for the Lyapunov exponent and the butterfly velocity. These rest essentially on the existence of a (semi)classical horizon with an infinite redshift, thus any infalling wave behaves as a shock wave and leads to a universal response. It seems that, despite the string/BH complementarity, a perturbative tree-level string calculation just cannot describe the transition to a black hole. One reason might be that we work in flat space background. This poses another important task for the future: will we get different results if we allow at least for weak gravity, taking it into account through perturbative vertex corrections? This question is in principle approachable and is another task for the future. Finally, one may try to model the dynamics at the transition point, which was studied in a number of recent works \cite{Chen:2020tes,Chen:2021Witten,Chen:2021BHstringsLargeD}.

\acknowledgments

We are grateful to Matthew Dodelson, David Berenstein and Dmitry Ageev for stimulating discussions. The authors acknowledge funding provided by the Institute of Physics Belgrade, through the grant by Ministry of Science, Technological Development, and Innovations of the Republic of Serbia. M.~\v{C}. would like to acknowledge the Steklov Mathematical Institute (Moscow), International Center for Theoretical Physics (Trieste) and Mainz Institute for Theoretical Physics (MITP) of the Cluster of Excellence PRISMA+ (Project ID 39083149) for hospitality and the opportunity to present and discuss this work.

\appendix

\bibliography{stringsbib.bib}

\end{document}